\documentclass{article}
\usepackage{latexsym,amsmath,amssymb,amsbsy,amstext,amscd,amsfonts}
\usepackage{graphics,graphicx}
\usepackage{epsfig} 
\usepackage{epstopdf}
\usepackage[margin=1in]{geometry}
\usepackage{subcaption}
\usepackage{color}
\usepackage{tabularx}
\usepackage{multirow}
\usepackage{booktabs}
\usepackage{float}
\usepackage{adjustbox}
\usepackage{mathrsfs}
\usepackage{cite}

\title{Transmission conditions for thin curvilinear close to circular heat-resistant interphases in composite ceramics}

\author{Daria Andreeva $^{(1,2)}$, Wiktoria Miszuris $^{(1,3)}$, Alexander Zagnetko$^{(1)}$,
\\
{\it $^{(1)}$ Department of Mathematics,
Aberystwyth University, }
\\ {\it Ceredigion SY23 3BZ, Wales, UK,}
\\{\it $^{(2)}$ Enginsoft Spa, }
\\ {\it Via della Stazione 27, fraz. Mattarello, 38123 Trento, Italy}
\\{\it $^{(3)}$ Eurotech Sp. z o.o., }
\\ {\it ul. Wojska Polskiego 3, 39-300 Mielec, Poland}}

\begin{document}

\maketitle

\begin{abstract}
This paper considers the problem of heat transfer in a composite ceramic material where the structural elements are bonded to the matrix via a thin heat resistant adhesive layer. The layer has the form of a circular ring or close to it. Using an asymptotic approach, the interphase is modeled by an infinitesimal imperfect interface,  preserving the main features of the temperature fields around the interphase, and allowing a significant simplification where FEM analysis is concerned. The nonlinear transmission conditions that accompany such an imperfect interface are evaluated, and their accuracy is verified by means of dedicated analytical examples as well as carefully designed FEM simulations. The interphases of various geometries are analysed, with an emphasis on the influence of the curvature of their boundaries on the accuracy of the evaluated conditions. Numerical results demonstrate the benefits of the approach and its limitations.
\end{abstract}

\section{Introduction}
The complex structure of composite ceramics with thin interphases enhances their thermal and mechanical properties \cite{2008, 2009, 2013, Mityakov}, which explains the wide use of such composites in modern technology. The presence of the thin interphase structures can sometimes present a challenge when modelling using finite element methods, leading to inaccuracies or numerical instabilities in the computations, which are clearly not acceptable, especially in such safety-conscious areas as automotive or aerospace technology \cite{FEM2008, 2013}.

One of the efficient approaches of tackling this problem consists of replacing the interphase in the numerical models with an infinitesimally thin (zero thickness) object, called {\it imperfect interface} and described by the specific transmission conditions \cite{2008,2009,2013,FEMver2008}. This method works effectively in various physical settings. In the context of elastic and acoustic scattering problems, it was used by B\"{o}vik \cite{Chalmers} in various cases, where the thin interphase was replaced by specially evaluated conditions dependent on the surrounding media, which could be fluid, viscous fluid or elastic material. In \cite{Hashin2001}, the imperfect interface conditions replacing the thin spherical interphase were derived in terms of the electrical conductivity, and were also directly applicable in the settings of thermal conductivity and magnetic fields. An arbitrary three-dimensional interphase, in the setting of dynamic elasticity and the transient heat problem, was considered in \cite{Ben2006}, which also contains a generalisation of the results from \cite{Chalmers} to the anisotropic case.

The replacement of the interphase with an imperfect interface is not always straightforward. In their paper \cite{Ben2010}, Benveniste and Berdychevsky give a comparison of two models in the context of elasticity: one in which the interphase is replaced by interface conditions, and a second in which the geometry is left unaltered and the interphase is simulated by conditions corresponding to each boundary. In \cite{Ben2012}, the ideas of \cite{Ben2006, Ben2010} were generalised in the case of the thin interphase having a non-constant conductivity.

In the works of Lebon and Rizzoni \cite{Rizzoni2010,Rizzoni2011,Rizzoni2012,Rizzoni2013}, a two-level model is introduced in such a manner that the interphase is replaced by a perfect interface at the first level, and by an imperfect interface at the second level, the latter being considered a correction of the leading solution. In \cite{Rizzoni2010}, they consider an adhesive joint from the point of view of displacement, while \cite{Rizzoni2011} extends the work to the anisotropic case. The method is then applied to a thin interphase with a mismatch strain in \cite{Rizzoni2012}, while \cite{Rizzoni2013} deals with a curved thin elastic interphase.

The interfacial/surface properties may change the effective properties of materials at the micro- or nano-scale (see the recent review in \cite{Erem_1}). An analysis of interface boundary conditions in the case of thermoelastic thin structures, and their influence on the deformation of shells and plates undergoing phase transformations, was performed in \cite{Erem_2} and \cite{Erem_3}. In the latter, in particular, the evolution of a circular interface under external loading was analysed.

All the aforementioned examples dealt with linear interphases of different physical natures. This method can also be effectively applied in the case of elastoplastic thin interphases \cite{Mish_1,Mish_2}. Here, the transformation to the respective imperfect interface models is far less trivial and requires several assumptions, as well as evaluation dependent on the yield condition \cite{Mish_3, Mish_4, Bord_1}.

The transmission conditions derived in this work will be used in the setting of heat or mass transfer (see an example of the latter described in \cite{Sanna}). Heat transfer problems for interphases with very flat (rectangular) surfaces and constant heat conductivity have been solved by Mishuris et al. in \cite{2008}, under the assumption that the temperature distribution within the interphase is monotonic. This result was later verified using FEM in \cite{FEMver2008}, and extended in \cite{2009} to the general case, with no assumptions on the monotonicity of the temperature.

Universal transmission conditions that do not require additional verification of specific conditions are given in \cite{2013}, in which both monotonic and non-monotonic temperature were considered, and the possible dependence of heat conductivity on temperature was taken into account. Also, \cite{FEM2008} gives a newly formulated FEM for thin interphases, making use of the ideas described in deriving the transmission conditions, and \cite{Mish_5} provides a classification and investigation of the edge effects that occur in thin interphases in the context of elasticity.

The case described in this paper is, essentially, the nonlinear problem of heat transfer in composites with thin layers that are represented as domains bounded by two smooth closed curves of circular or close to circular shape. The reason for concentrating on such geometries is simply that this is the most common shape of inclusions in ceramic composites. The material of the interphase has low conductivity compared to that of its two surrounding ceramic layers (which are not necessarily of the same material). Section 2 gives a detailed problem formulation, as well as some preliminary steps (the transition to curvilinear coordinates and rescaling of the interphase). The process for evaluating the transmission conditions is presented in Section 3, while Section 4 supports the obtained results with numerical examples. Finally, Section 5 concludes this article and sets the scope of further work on this problem.

\section{Problem formulation}
We consider the case of a thin interphase placed between two media. Within the interphase, the heat transfer equation
\begin{equation}\label{Heat}
\nabla \cdot (k\nabla T) + Q = c \rho \dfrac{\partial T}{\partial t},
\end{equation}
is satisfied, where $T(x,y)$ is the unknown temperature, $Q(T,x,y)$ the thermal source, $k(T,x,y)$ the thermal conductivity, $c$ the heat capacity and $\rho$ the density of the material. We note that the assumption is that $Q$ does not change its sign within the interphase, as it would otherwise make little sense to have multiple heat sources and sinks in an interphase so thin.
\begin{figure}[H]
\begin{center}
\includegraphics[scale=0.5]{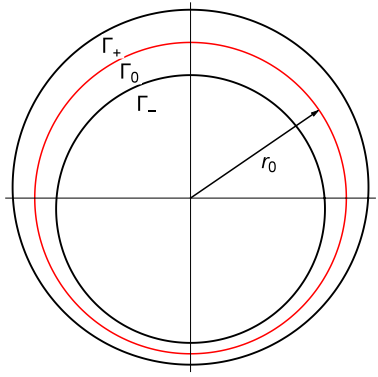}\caption{Thin, almost circular interphase}\label{Fig.0}
\end{center}
\end{figure}
Transmission conditions along the boundaries of the interphase $\Gamma_{\pm}$ between the corresponding materials are written in the form
\begin{equation}
\begin{array}{rclcrcl}
\left[T\right]|_{\Gamma_{+}} &=& 0, &\quad& \left[T\right]|_{\Gamma_{-}} &=& 0, \\[2mm]
\left[\textbf{n} \textbf{q}\right]|_{\Gamma_{+}} &=& 0, &\quad& \left[\textbf{n} \textbf{q}\right]|_{\Gamma_{-}} &=& 0,
\end{array}
\end{equation}
where $\textbf{n}$ is the vector normal to the surface, and $\textbf{q}$ is the heat flux.

The heat flux is defined according to Fourier's law,
\begin{equation}\label{HeatFlux}
\textbf{q} = -k(T) \nabla T.
\end{equation}

We should mention at this point that analogous transmission conditions can be derived when considering a problem of mass, rather than heat, transfer by using the first and second Fick's laws instead of (\ref{HeatFlux}) and (\ref{Heat}) respectively. We shall, however, proceed by using only the heat transfer terminology.

As the temperature varies for all three media, it is described by a distinct function within each region:
$T_{+}(x, y, t)$ from the outer edge of the interphase to the outside of the outer border $\Gamma_{+}$ of the interphase; $T_{-}(x, y, t)$ in the region bounded by the inner border $\Gamma_{-}$ of the interphase; $T(x, y, t)$ in the interphase itself.

Each material is also characterised by its thermal conductivity, denoted by $k_{+}, k, k_{-}$ respectively, where $k \ll k_{\pm}$ and, generally speaking, $k_{+} \neq k_{-}$.

\subsection{Transition to polar coordinates and rescaling of the interphase}
Taking the form of the interphase into account, it is convenient to switch to polar coordinates $(r, \phi)$. The heat transfer equation (\ref{Heat}) then transforms into
\begin{equation}
\label{HeatTransformedPolar}
\frac{1}{r}\frac{\partial}{\partial r}
\left(kr\frac{\partial T}{\partial r}\right)+\frac{1}{r^2}\frac{\partial}{\partial \phi}
\left(k\frac{\partial T}{\partial \phi}\right)+Q = c \rho \frac{\partial T}{\partial t},
\end{equation}
and the transmission conditions along the interphase in polar coordinates are given by
\begin{eqnarray} \label{TConditionPolar}
T_{\pm} - T(r_{\pm}, \phi, t) &=& 0,  \\
\label{QConditionPolar}
q_{\pm} +
n_r^\pm k \frac{\partial}{\partial r} T(r_{\pm}, \phi, t) +
n_{\phi}^\pm k \frac{1}{r_\pm} \frac{\partial}{\partial \phi} T(r_{\pm}, \phi, t) &=& 0.
\end{eqnarray}

Within these formulae, and henceforth, $q_+, q$ and $q_-$ stand for the normal component of the heat flux $\textbf{q}$, with respect to the relevant boundary,
within the material regions, and defined by the polar coordinates $(r_+, \phi)$, $(r, \phi)$ and $(r_-, \phi)$ respectively.

Here $r=r_{\pm}(\phi)$ stands for the equations of the boundaries $\Gamma_{\pm}$ in polar coordinates, and the normal vectors to the boundaries in polar coordinates are defined by
\begin{equation}\label{NormalVector}
\mathbf{n_{\pm}} = [n_{r}^{\pm}, n_{\phi}^{\pm}] = \frac{1}{\sqrt{r_{\pm}^2(\phi) + (r'_{\pm}(\phi))^2}} \Big[r_{\pm}(\phi), r'_{\pm}(\phi)\Big].
\end{equation}

We also introduce $r = r_0(\phi)$, denoting the center line $\Gamma$ between the two interfacial boundaries $\Gamma_{+}$ and $\Gamma_{-}$. Due to the shape chosen for the interphase, the center line is constant in polar coordinates, i.e. $r(\phi)\equiv r_0$.

The thin interphase for this kind of problem is commonly rescaled (\cite{2008, 2013}) utilising:
\begin{equation}\label{Xi}
\xi = \frac{r - r_0}{\varepsilon \widetilde{h}},\quad \xi\in(-1/2,1/2),
\end{equation}
instead of the coordinate $r \in (r_{-}, r_{+}) = \left(r_0 - h/2, r_0 + h/2\right)$,
while $h$ is the width of the thin interphase:
\begin{equation}
h = h(\phi) = r_{+}(\phi) - r_{-}(\phi)=\varepsilon \widetilde{h}(\phi),
\end{equation}
where $\varepsilon \ll 1$. In the rescaling, the second coordinate remains unchanged $\phi \in [0, 2\pi)$.

We assume that the thermal conductivity $k$ of the interphase is much smaller than those of the surrounding media,$(k_{\pm})$, and should also be similarly rescaled:
\begin{equation}
k(T, r, \phi) = \varepsilon \widetilde{k}(T, \xi, \phi),
\end{equation}
with the same small parameter $\varepsilon \ll 1$.
Finally, we introduce the following notation
\begin{equation}
Q = \frac{1}{\varepsilon} \widetilde{Q},
\end{equation}
indicating that we deal with the essential heat source/sink acting inside the interphase.

After rescaling, the temperature is described by the function $\widetilde{T}$
\begin{equation}
T(r, \phi, t) = \widetilde{T}(\xi, \phi, t).
\end{equation}

As a consequence of transferring to the new variables ($\xi,\phi$), equations (\ref{HeatTransformedPolar}) takes the form:
\begin{eqnarray}
\frac{1}{\varepsilon} \widetilde{Q} + \frac{1}{\varepsilon \tilde{h}^2} \frac{\partial}{\partial \xi}\Big(\tilde{k} \frac{\partial \tilde T}{\partial \xi}\Big) + \frac{\tilde{k}}{\tilde{h}(r_0+\varepsilon\tilde{h}\xi)} \frac{\partial \tilde T}{\partial \xi} &+& \label{HeatRescaled} \\
\frac{1}{(r_0+\varepsilon\tilde{h}\xi)^2} \Big(
\frac{r_0'+\varepsilon\tilde{h}'\xi}{\varepsilon \tilde{h}}  \frac{\partial}{\partial \xi} -  \frac{\partial}{\partial \phi}
\Big) \tilde{k} \Big(\frac{r_0'+\varepsilon\tilde{h}'}{\tilde{h}} \frac{\partial \widetilde{T}}{\partial \xi} - \varepsilon \frac{\partial \widetilde{T}}{\partial \phi}
\Big)&=& c \rho \frac{\partial \widetilde{T}}{\partial t}. \nonumber
\end{eqnarray}

For the sake of brevity, from hereon we shall write $\widetilde{h}$ in place of $\widetilde{h}(\phi)$, and $r_0$ in place of $r_0(\phi)$.

The transmission conditions (\ref{TConditionPolar}) and (\ref{QConditionPolar}), after rescaling, transform into
\begin{eqnarray}
T_{\pm} - \widetilde{T}\Big(\pm \frac{1}{2}, \phi, t\Big) &= 0, \label{TConditionRescaled} \\
q_{\pm} +
n_\xi^\pm \frac{\widetilde{k}}{\widetilde{h}} \frac{\partial \widetilde{T}}{\partial \xi} +
\frac{ n_\phi^\pm \widetilde{k}}{r_0(\phi) \pm \frac{1}{2} \varepsilon \widetilde{h}(\phi)}\Big(\frac{r_0' \pm \varepsilon \xi \tilde h'}{\tilde h} \frac{\partial \widetilde{T}}{\partial \xi} -  \varepsilon \frac{\partial \widetilde{T}}{\partial \phi}\Big)
\Big\vert_{\big(\pm \frac{1}{2}, \phi, t\big)}
 &= 0. \label{QConditionRescaled}
\end{eqnarray}

\section{Evaluation of the transmission conditions}
We seek a solution for the temperature $\widetilde{T}$ within the interphase, in the form of an asymptotic expansion,
\begin{equation}\label{TExpansion}
\widetilde{T} = \widetilde{T}_0(\xi, \phi, t) + \varepsilon \widetilde{T}_1(\xi, \phi, t) + \varepsilon^2 \widetilde{T}_2(\xi, \phi, t) + ...
\end{equation}

Substituting this expansion into (\ref{HeatRescaled}), we obtain a set of boundary value problems for the consecutive terms $\widetilde{T}_j(\xi, \phi, t)$.
We note that only the first problem is nonlinear
\begin{equation}
\frac{1}{\widetilde{h}^2}\frac{\partial }{\partial \xi} \left(\widetilde{k}\frac{\partial \widetilde{T}_0}{\partial \xi}\right)
 + \widetilde{Q} =0.
\label{BVP}
\end{equation}

while the second and implied further problems are linear. In the following we restrict ourselves to the leading terms for temperature and heat flux.

Finally, in order to write boundary conditions for this equation along the boundaries $\Gamma_+$ and $\Gamma_-$ we also need to
expand the normal vector in powers of $\varepsilon$ to obtain the estimate:
\begin{equation}\label{NormalVectorExpansion}
\mathbf{n_\pm} = [1,0] + O(\varepsilon).
\end{equation}

After insertion of the expansions (\ref{TExpansion}) and (\ref{NormalVectorExpansion})  the transmission conditions (\ref{TConditionRescaled}) and (\ref{QConditionRescaled}), we can formulate the boundary value problem for equation (\ref{BVP}) with the boundary conditions:
\begin{eqnarray}
T_{\pm} - \widetilde{T}_0\Big(\pm \frac{1}{2}, \phi, t\Big) &=& 0, \label{BoundaryTemperature} \\
q_{\pm} + \frac{1}{\widetilde{h}} \widetilde{k}\Big(\widetilde{T}_0, \pm \frac{1}{2}, \phi\Big) \frac{\partial \widetilde{T}_0}{\partial \xi} \Big\vert_{\big(\pm \frac{1}{2}, \phi, t\big)} &=& 0. \label{BoundaryHeatFlux}
\end{eqnarray}

We note that the first component of the heat flux at a point $(\xi, \phi)$ inside the rescaled interphase can be estimated by
\begin{equation}\label{RescaledHeatFlux}
\widetilde{q}_\xi = - \frac{1}{\widetilde{h}} \widetilde{k}\Big(\widetilde{T}_0(\xi, \phi, t), \xi, \phi\Big)
\frac{\partial}{\partial \xi} \widetilde{T}_0(\xi, \phi, t) + O(\varepsilon).
\end{equation}

\subsection{General algorithm of evaluation of the transmission conditions}
In order to evaluate the transmission conditions, we need to integrate the differential equation (\ref{BVP}) with respect to variable $\xi$, obtaining the integral
\begin{equation}
\mathscr{F}(T, \xi, C_0, C_1) = 0.
\end{equation}
Differentiating this integral provides another equation,
\begin{equation}
- \frac{\widetilde{k}}{\widetilde{h}} \frac{\partial}{\partial \xi}\mathscr{F}(T, \xi, C_0, C_1) =
\widetilde{q}_\xi \frac{\partial}{\partial T}\mathscr{F}(T, \xi, C_0, C_1).
\end{equation}

Then, by substituting the values $\xi = -\frac{1}{2}$, $T = T_-$, $q=q_-$, we find the integration constants $C_0$, $C_1$. Subsequent substitution of $\xi = \frac{1}{2}$, $T = T_+$, $q=q_+$ reveals two necessary solvability conditions,
\begin{equation}
\begin{array}{rcl}
F_1(T_+, T_-, q_+, q_-) &=& 0, \\
F_2(T_+, T_-, q_+, q_-) &=& 0.
\end{array}
\label{GeneralConditions}
\end{equation}
In this manner, we obtain the transmission conditions for an arbitrary case. In some special cases, represented in detail in Appendix A, conditions (\ref{GeneralConditions}) can be written explicitly. We note that, in case the temperature within the interphase is non-monotonic, the derived conditions also provide formulae for finding its maximum/minimum point (see Appendix A.2 and A.3).

\subsubsection{Transmission conditions in the linear case}
We first consider the simplest case, when the material parameters are temperature-independent, or, in other words,
\begin{equation*}
\widetilde{Q} = \widetilde{Q}(\xi,\phi,t), \quad \widetilde{k} = \widetilde{k}(\xi,\phi,t).
\end{equation*}
In this particular situation, the transmission conditions (\ref{GeneralConditions}) take the linear forms
\begin{equation}
\begin{array}{rcl}
q_+- q_- &=& A_1^{(1)}(\phi,t), \\
T_+- T_- &=& B_1^{(1)}(\phi,t)q_-+B_2^{(1)}(\phi,t).
\end{array}
\label{Conditions_Case_1}
\end{equation}
Their explicit formulae can be found in Appendix A.1.

\subsubsection{Transmission conditions in case of the thermal conductivity being a function of temperature}
Let us now assume that
\begin{equation}
\widetilde{Q} = \widetilde{Q}(\xi,\phi,t), \quad \widetilde{k} = \widetilde{k}(T,\phi,t)m(\xi,\phi,t).
\end{equation}
Although this is not a general case in terms of conductivity, this representation is sufficient, as it includes also the case of $k=k(T,\xi,\phi,t)$.

If the temperature is distributed monotonically within the interphase, then (\ref{GeneralConditions}) can be transformed to the view
\begin{equation}
\begin{array}{rcl}
q_+- q_- &=& A_1^{(2)}(\phi,t), \\[2mm]
B_0^{(2)}(T_+,T_-,\phi,t) &=& B_1^{(2)}(\phi,t)q_-+B_2^{(2)}(\phi,t),
\end{array}
\label{Conditions_Case_2}
\end{equation}
where $A_1^{(2)} \equiv A_1^{(1)}$, i.e. the first transmission condition and process of obtaining it is also exactly the same as in the previous case.
In the general case, when we discard the monotonicity assumptions, the conditions take more complex forms (see Appendix A.2). They can be written as
\begin{equation}
\begin{array}{rcl}
A_+^{(2)}(q_+,\phi,t) &=& A_-^{(2)}(q_-,\phi,t), \\[2mm]
B_+^{(2)}(q_+,T_+,\phi,t) &=& B_-^{(2)}(q_-,T_-,\phi,t).
\end{array}
\label{Conditions_Case_2_NM}
\end{equation}

\subsubsection{Transmission conditions for temperature-dependent sources and material properties}
We now assume that both the thermal conductivity and heat source depend on the temperature, but not the coordinates
\begin{equation}
\widetilde{Q} = \widetilde{Q}(T,\phi,t), \quad \widetilde{k} = \widetilde{k}(T,\phi,t),
\end{equation}
and obtain the transmission conditions in the form
\begin{equation}
\begin{array}{rcl}
q_+^2 - q_-^2 &=& A_1^{(3)}(T_+,T_-,\phi,t), \\[2mm]
B_{1 \pm}^{(3)}(q_\pm, T_\pm, T_\mp,\phi,t) &=& B_0^{(3)}(\phi,t),
\end{array}
\label{Conditions_Case_3}
\end{equation}
if the temperature is assumed to be monotonic, and
\begin{equation}
\begin{array}{rcl}
A_+^{(3)}(q_+,T_+,\phi,t) &=& A_-^{(3)}(q_-,T_-,\phi,t), \\[2mm]
B_{+*}^{(3)}(q_+,T_+,\phi,t)+B_{-*}^{(3)}(q_-,T_-,\phi,t) &=& B_0^{(3)}(\phi,t).
\end{array}
\label{Conditions_Case_3_NM}
\end{equation}
otherwise. Appendix A.3 provides the explicit expressions.

\section{Numerical examples}
\begin{figure}[H]
\begin{center}
\begin{minipage}{0.3\linewidth}
\includegraphics[scale=0.4]{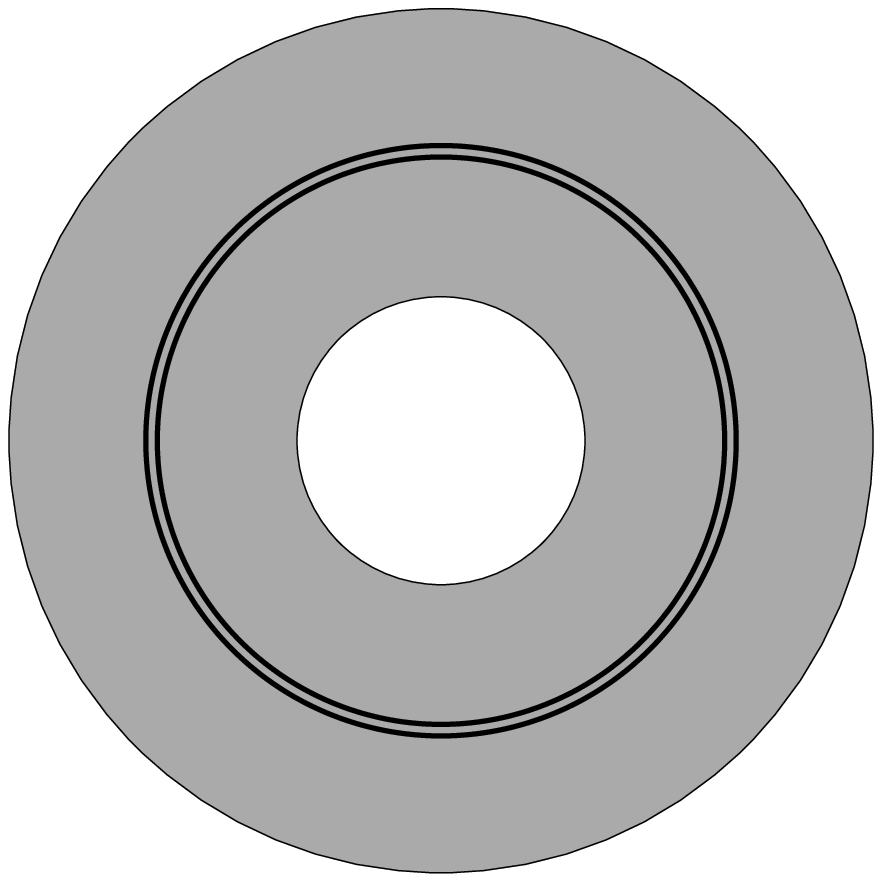}\subcaption{\quad }
\end{minipage}
\begin{minipage}{0.3\linewidth}
\includegraphics[scale=0.4]{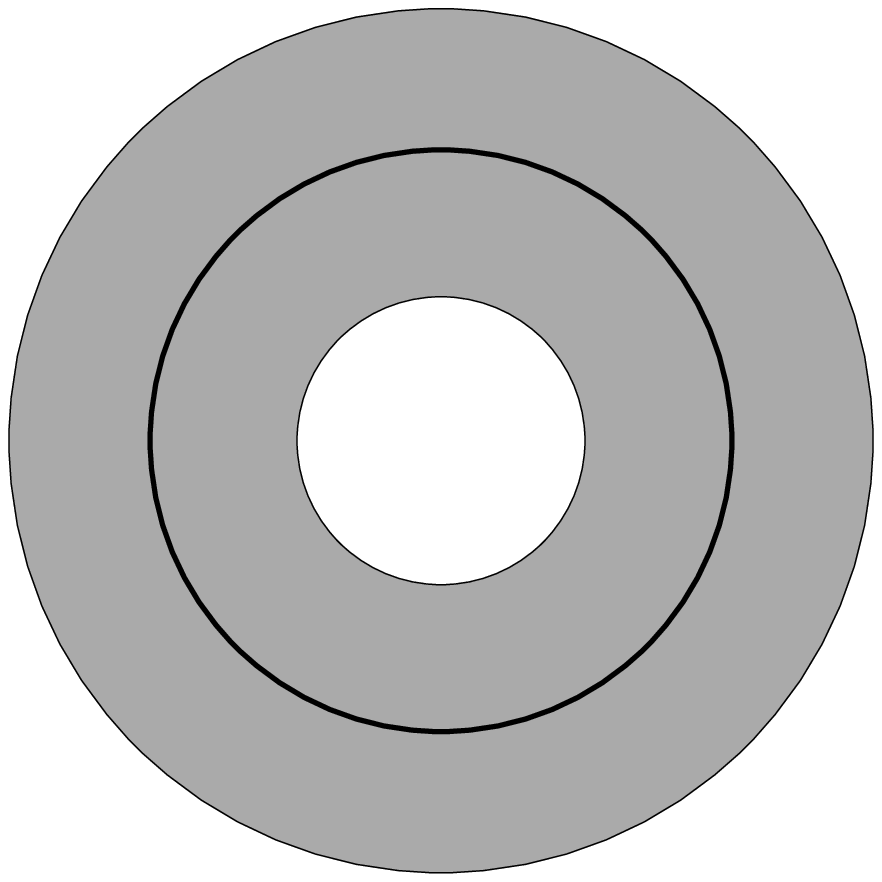}\subcaption{\quad }
\end{minipage}
\end{center}
\begin{center}
\begin{minipage}{0.3\linewidth}
\includegraphics[scale=0.4]{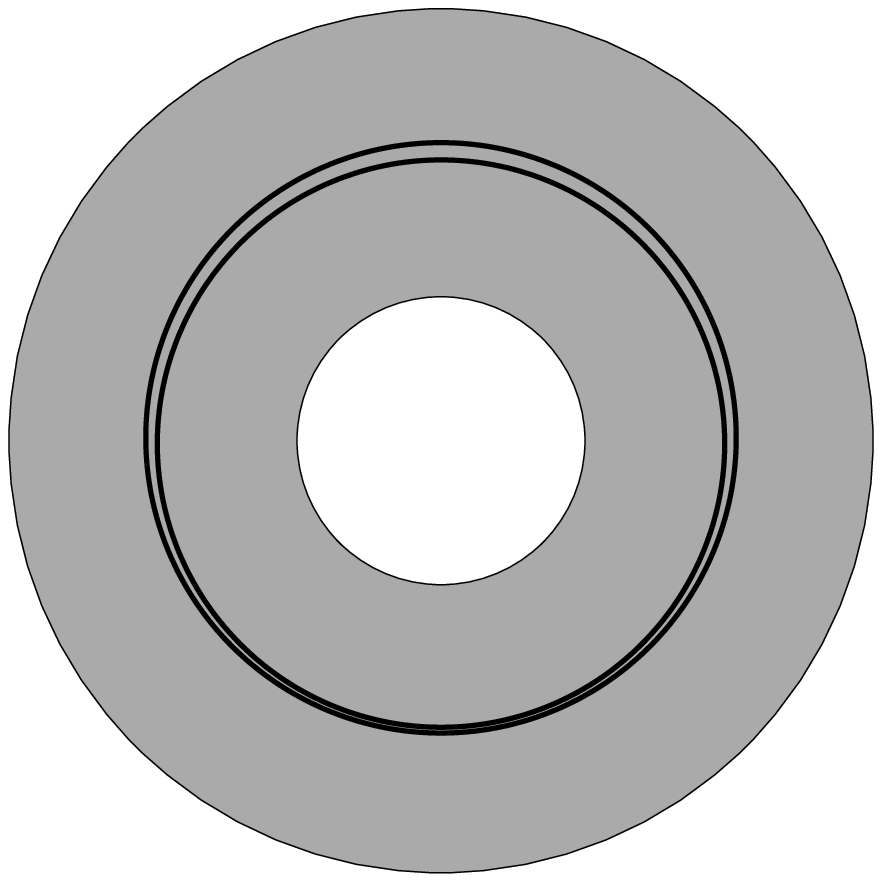}\subcaption{\quad }
\end{minipage}
\begin{minipage}{0.3\linewidth}
\includegraphics[scale=0.4]{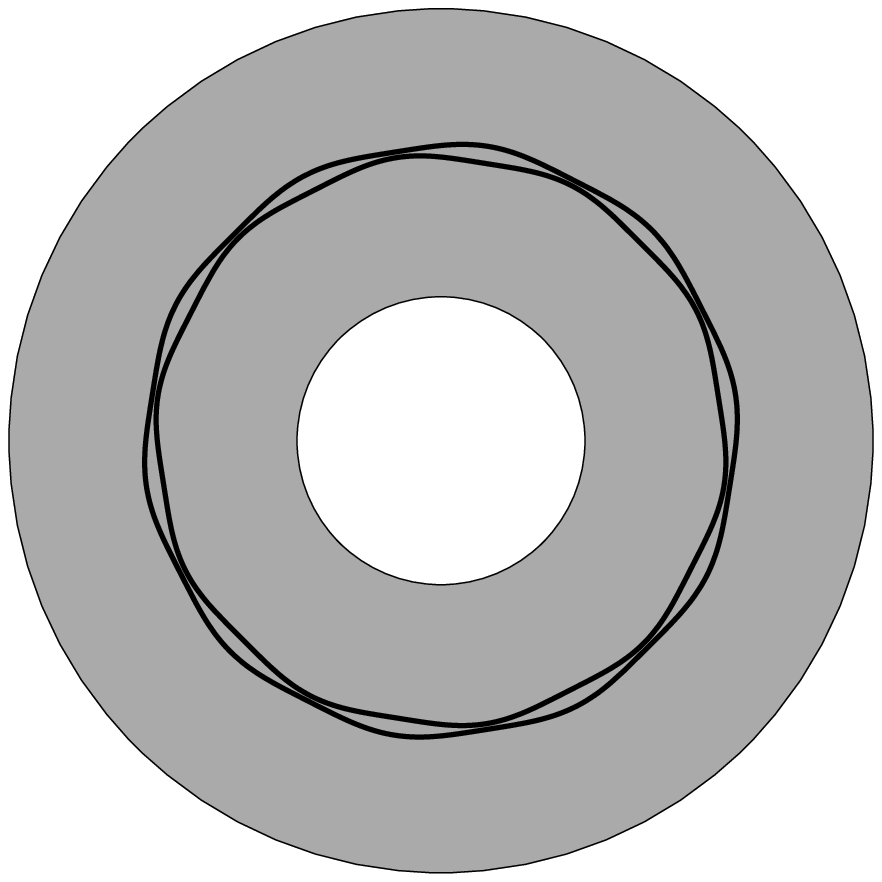}\subcaption{\quad }
\end{minipage}
\begin{minipage}{0.3\linewidth}
\includegraphics[scale=0.4]{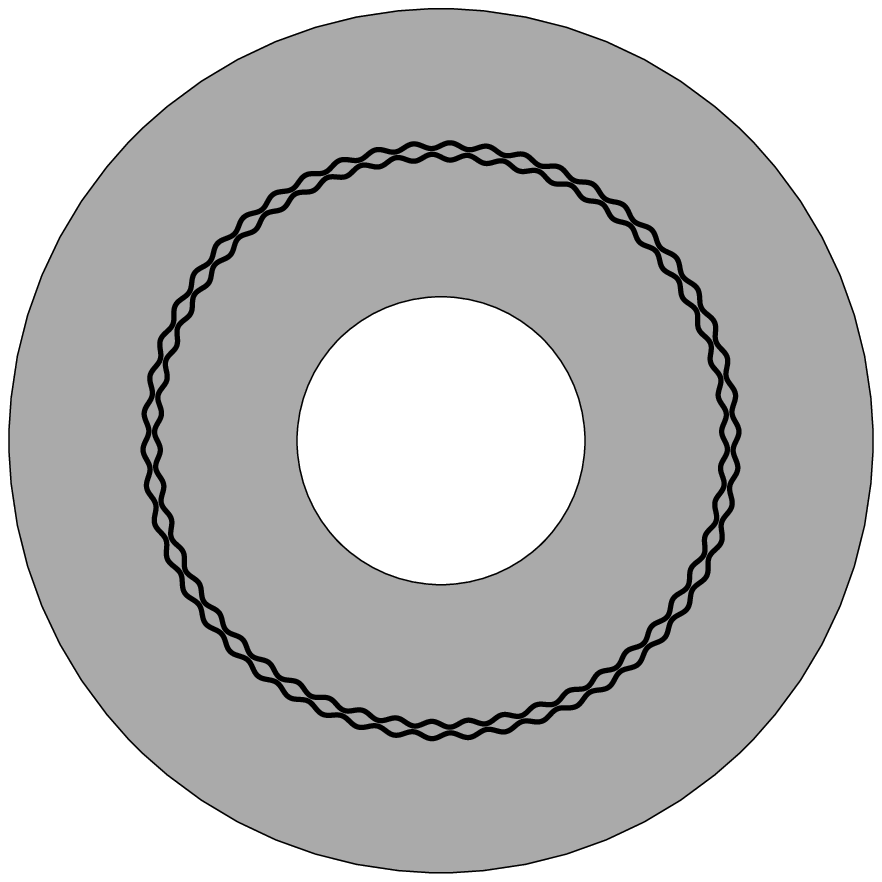}\subcaption{\quad }
\end{minipage}
\end{center}
\caption{Domains with circular thin interphase (a) of the constant thickness corresponding to $n=0$ (see (\ref{shapes})); with almost circular interphases of various curvature (c-e) (here $n=1,10,50$); the problem with the imperfect interface of zero thickness (b), where the shape of the thin original interphase is accounted for in the transmission conditions (\ref{GeneralConditions}).}
\label{Fig.1}
\end{figure}

The key purpose of this section is to demonstrate how the previously evaluated transmission conditions can lead to an accurate approximation of the exact solutions. To verify the numerical approximations against analytically obtained solutions, we considered the temperature distribution within an inhomogeneous cylindrical domain with an interphase of circular or almost circular shape (see Fig.\ref{Fig.1}). The interphase's boundaries are described by the equations
\begin{equation}
\label{shapes}
r_\pm(\phi)=1 \pm 0.005(2+\sin n \phi),
\end{equation}
where $n=0, 1, 10, 50$. Note that $n=0$ corresponds to the case of a circular interphase.

The problem is then considered in two formulations: the original problem with the thin interphase, and the simplified formulation where the thin interphase is replaced by an imperfect interface of zero thickness ($r=r_0=r_-=r_+$).

\subsection{Case of circular interphase}
We first assume that the interphase is in the form of a thin ring.
Such a simple shape makes it possible to analytically find solutions of both problems, involving the interphase and imperfect interface, reducing them to one-dimensional problems.
The circular hole inside the cylinder allows to create both monotonic and non-monotonic temperature distributions within the interphase by manipulations of the boundary condition on the inner surface, and thus to verify both sets of transmission conditions. Precisely, we consider a dimensionless domain with outer (inner) boundary at $R = 1.5$ ($\widehat{r} = 0.5$) and the interphase boundaries $r_+ = 1.01, r_- = 1$. Thus $h = r_+ - r_- = 0.01$.

\subsubsection{Parameter values and normalisation}
The interphase has thermal conductivity $k_0 = 0.2$, while the thermal conductivities of the surrounding materials are assumed to be equal: $k_+ = k_- = 237$ (as in \cite{FEMver2008}).
The heat source is assumed to be $Q(T, r) = 50 (50 + T)$ and $Q(T, r) = 500 (50 + T)$ for the cases a monotonic and non-monotonic temperature distribution within the interphase respectively, while for all the problems analysed we set the temperature to $\widehat{T} = 295$ at the outer and to $T_0 = 300$ at the inner boundary.
Note that the small parameter in this case can be chosen as $\varepsilon=h=0.01$.

\subsubsection{Case of monotonic temperature in the interphase}
We solved this problem analytically, and numerically with the help of the FEM software COMSOL Multiphysics$^{\circledR}$ version 4.3b. Analytic solutions were found for both the original problem with the thin interphase and the approximate one with the imperfect interface. The exact formulae for both can be found in Appendix B.1. Here and further, when we speak about the solution to the simplified problem within the interphase, we mean the temperature distribution in the region geometrically corresponding to the interphase in the original problem, as the interphase itself has been replaced with the imperfect interface.

\begin{figure}[H]
\begin{minipage}[h]{0.48\linewidth}
\includegraphics[scale=0.47]{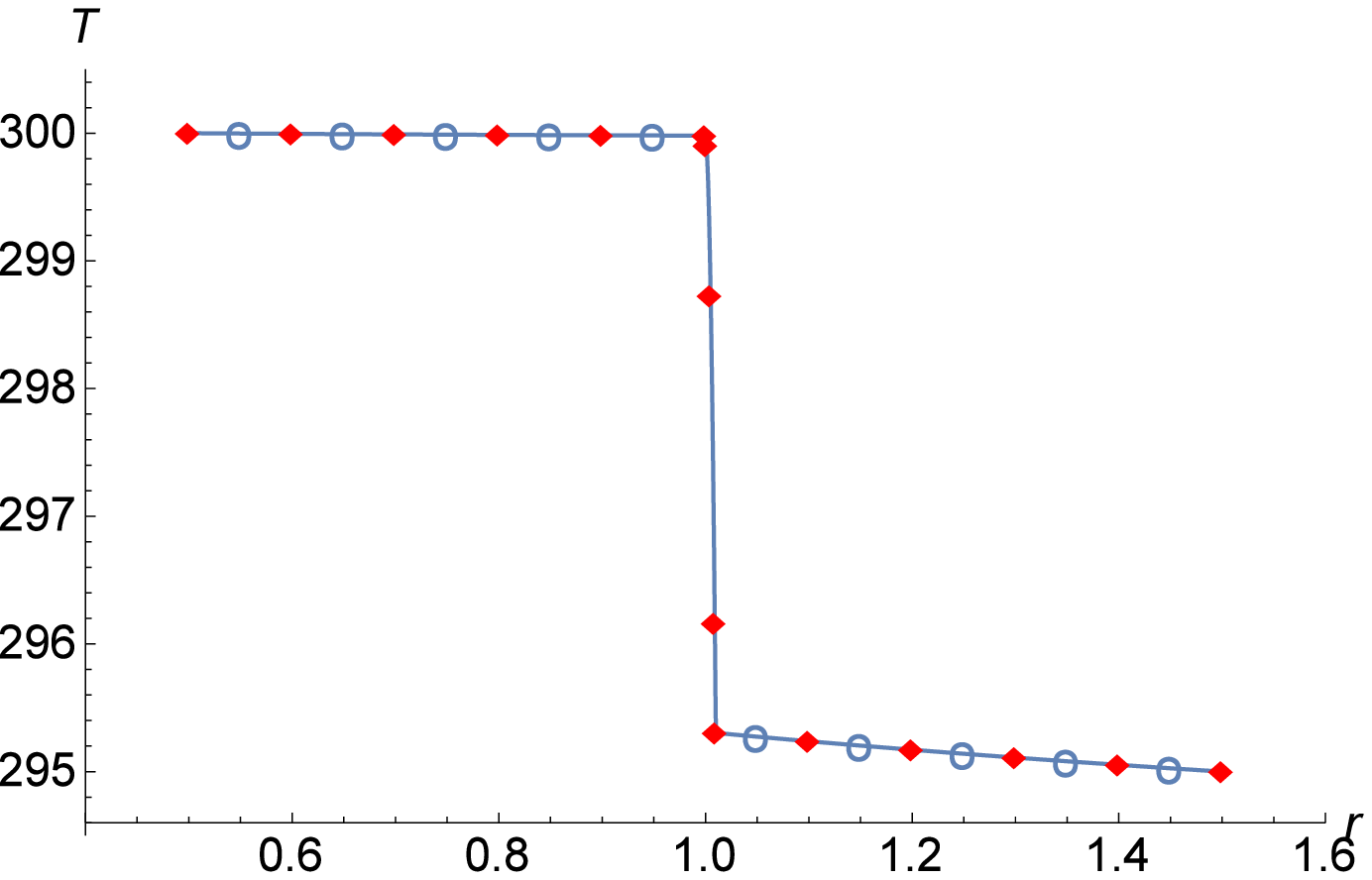}\subcaption{\quad }\label{Fig.2a}
\end{minipage} \hspace{0.01\linewidth}
\begin{minipage}[h]{0.48\linewidth}
\includegraphics[scale=0.7]{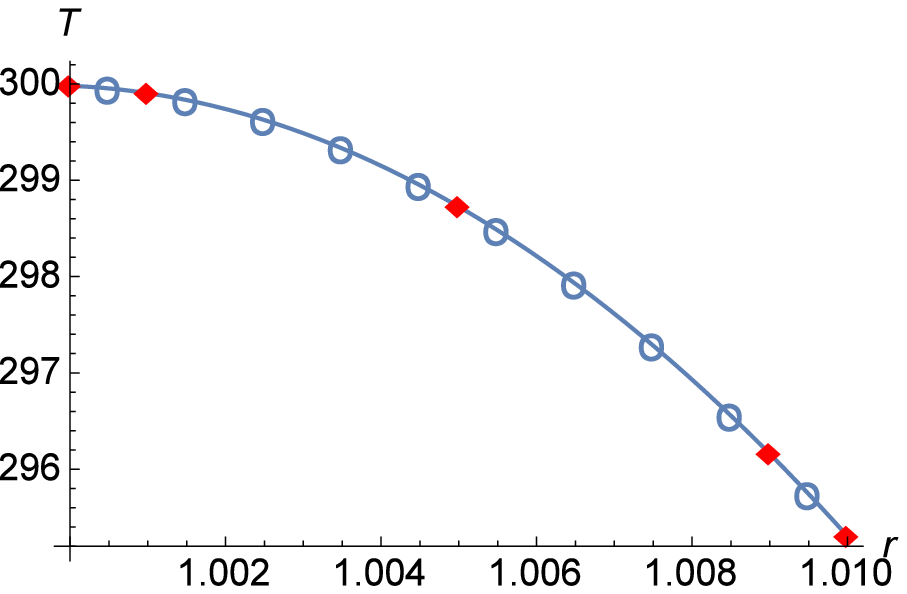}\subcaption{\quad }\label{Fig.2b}
\end{minipage}
\begin{minipage}[h]{0.48\linewidth}
\includegraphics[scale=0.47]{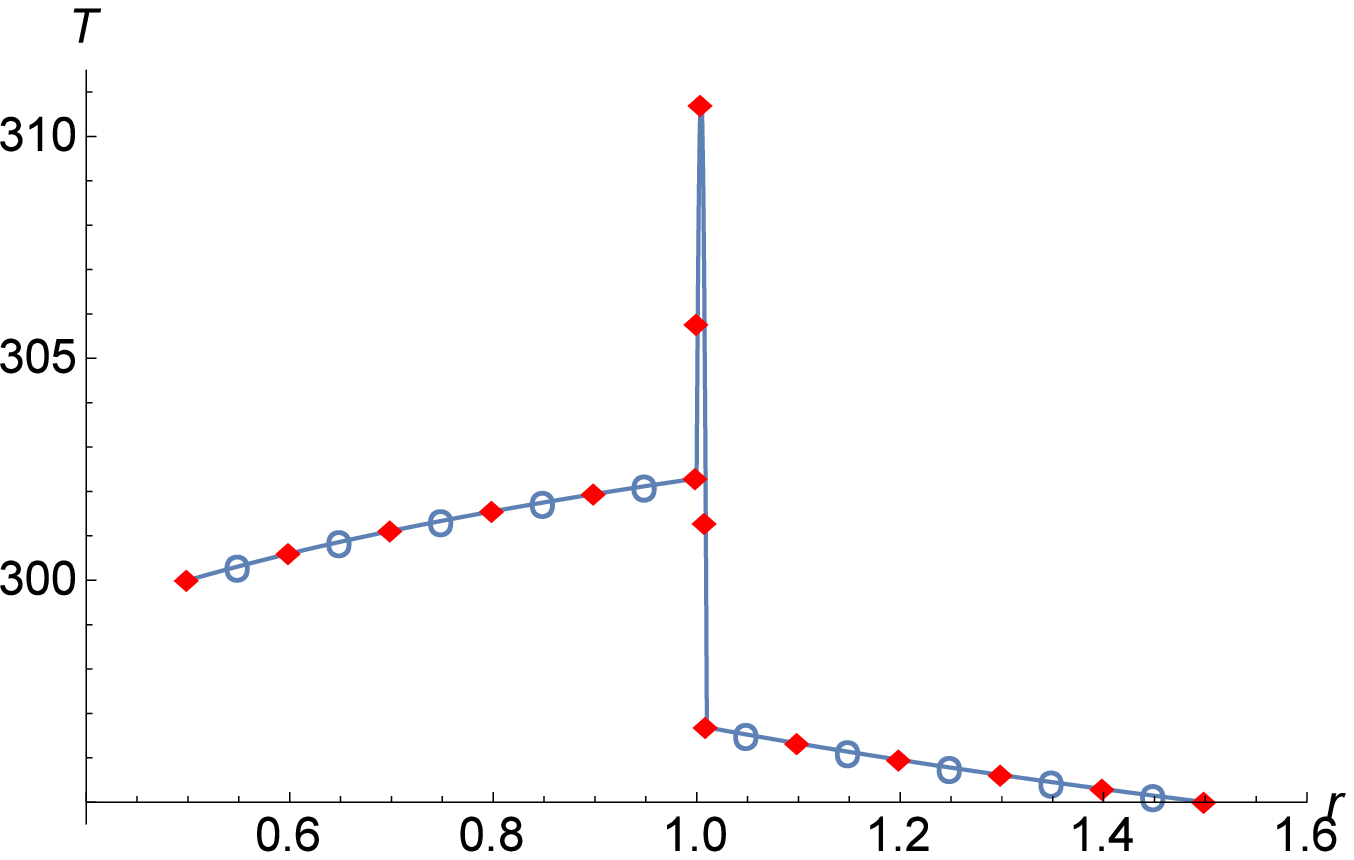}\subcaption{\quad }\label{Fig.4a}
\end{minipage} \hspace{0.01\linewidth}
\begin{minipage}[h]{0.48\linewidth}
\includegraphics[scale=0.7]{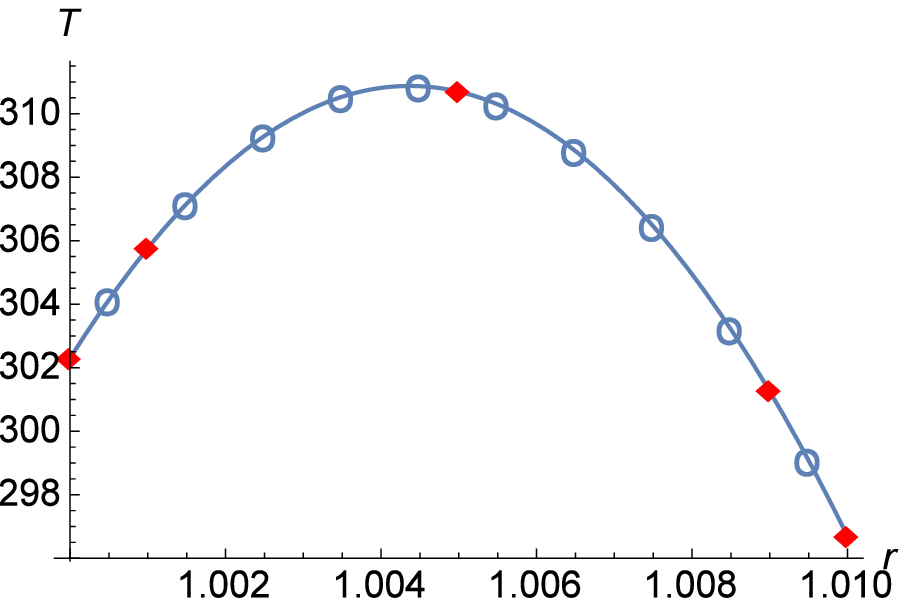}\subcaption{\quad }\label{Fig.4b}
\end{minipage}
\caption{The analytical solution (solid line), COMSOL solution (diamond markers) and the solution to the simplified problem with the imperfect interface (circular markers) for the cases: (a-b) -- monotonic temperature inside the interphase and (c-d) -- non-monotonic $T(r)$.}
\end{figure}

In Fig.~\ref{Fig.2a}, we present the analytically obtained solution (the solid line) and the values obtained in COMSOL (the diamond markers), both for the original problem with the thin circular interphase, and the approximation obtained by solving the simplified problem with transmission conditions (circular markers). We also include
a close-up of the temperature distribution in the interphase only in Fig.~\ref{Fig.2b}, where a perfect match, indistinguishable by sight in the original scale, can be observed.

The COMSOL results shown in Fig.~\ref{Fig.2a} and \ref{Fig.2b}, were obtained with the COMSOL option {\it `extremely fine mesh'}; the usage of the {\it `normal mesh'} option barely influences the results in this case. The levels of error are shown in Table.~\ref{TemperatureErrors}, which includes typical absolute errors, $\Delta$, in each material and inside the interphase, and their corresponding relative errors, $\delta$. In this table, $T_1$ stands for the exact solution of the approximate formulation of the problem with the transmission conditions, $T_2$ represents the numerical solution computed in COMSOL, and $r_1$ denotes the maximum point evaluated through the transmission conditions (which in this case coincides with the boundary at which the temperature is the highest).

\begin{table}[H]
\begin{center}
\begin{tabular}{l*{11}{c}}
 & $\Delta T_1$ & $\delta T_1$ & $\Delta T_2$ & $\delta T_2$ & $\Delta r_1$ & $\delta r_1$\\
\hline
M & 0.007 & $2.33*10^{-5}$ & 0.002 & $6.78*10^{-6}$ & $1.86*10^{-5}$ & $1.86*10^{-5}$ \\
\hline
N-M & 0.014 & $4.65*10^{-5}$ & 0.030 & $9.82*10^{-5}$ & $5*10^{-6}$ & $4.98*10^{-6}$
\end{tabular}
\caption{Errors for the exact solution of the simplified problem with the transmission condition ($T_1$), numerical (COMSOL) solution ($T_2$) and the evaluated maximum point ($r_1$), where the cases of monotonic and non-monotonic temperature are denoted by `M' and `N-M' respectively.}\label{TemperatureErrors}
\end{center}
\end{table}

\subsubsection{Case of non-monotonic temperature in the interphase}
We use the same approach in the second case, when the solution is not monotonic inside the interphase.
The explicit formulae for the analytical solutions are given in Appendix B.2. Similarly to the preceding case, the solutions are presented in Fig.~\ref{Fig.4a} (for the whole domain) and Fig.~\ref{Fig.4b} (for the interphase region only), and the accompanying errors, including those at the evaluated maximum point, are shown in Table.\ref{TemperatureErrors}.
Here, again, we speak of the COMSOL results obtained on the extremely fine mesh, as computations for the normal mesh gave only negligible difference.

Both cases demonstrated that the imperfect interface approach is extremely efficient and produced a relative error of the order $O(h^2)$ for the analysed thickness of the interphase ($h=0.01$). We note that decreasing the thickness makes it impossible to compute the numerical solution using COMSOL while the analytical solutions for both the original interphase problem and the approximate version using the imperfect interface give the same $O(h^2)$ order of accuracy.

Having checked the accuracy of the solutions computed by various methods, we proceed to analyse the accuracy of the transmission conditions themselves.
However, this has already been accomplished, although indirectly, by comparing the solution to the problem simplified using the imperfect transmission conditions with the solution to the original formulation.

\subsubsection{Verification of transmission conditions}
We can check the precision of the transmission conditions themselves by substituting the exact solution into the transmission conditions and evaluating the relative error between the left and right sides of the resulting equations. The respective errors are denoted by $\delta_1^{(m)}$ and $\delta_2^{(m)}$ for the first and the second conditions obtained under assumptions of monotonicity, and $\delta_1^{(nm)}$ and $\delta_2^{(nm)}$ without such assumptions. The explicit formulae for calculating the errors are given in Appendix C. The collected errors are shown in Table.~\ref{Verification1}. We can see that the conditions are precise in both cases, with errors of the orders $10^{-2}$ -- $10^{-4}$.
We note that $\delta_1^{(m)}=\delta_1^{(nm)}$, as the first transmission conditions, evaluated under assumptions of monotonic and non-monotonic temperature distributions, coincide.
\begin{table}[H]
\begin{center}
\begin{tabular}{l*{6}{c}}
relative errors & $\delta_1^{(m)}$ & $\delta_2^{(m)}$ & $\delta_2^{(nm)}$  \\
\hline
monotonic $T(r)$ & 0.007 & 0.003 & 0.005\\
\hline
non-monotonic $T(r)$ & 0.014 & $0.0008$ & $0.0005$\\
\end{tabular}
\caption{Precision of the transmission conditions, verified by substituting into them the exact solutions.}
\label{Verification1}
\end{center}
\end{table}

\subsubsection{Physical parameters along the boundaries of the interphase}
\begin{figure}[H]
\begin{minipage}[h]{0.48\linewidth}
\includegraphics[scale=0.6]{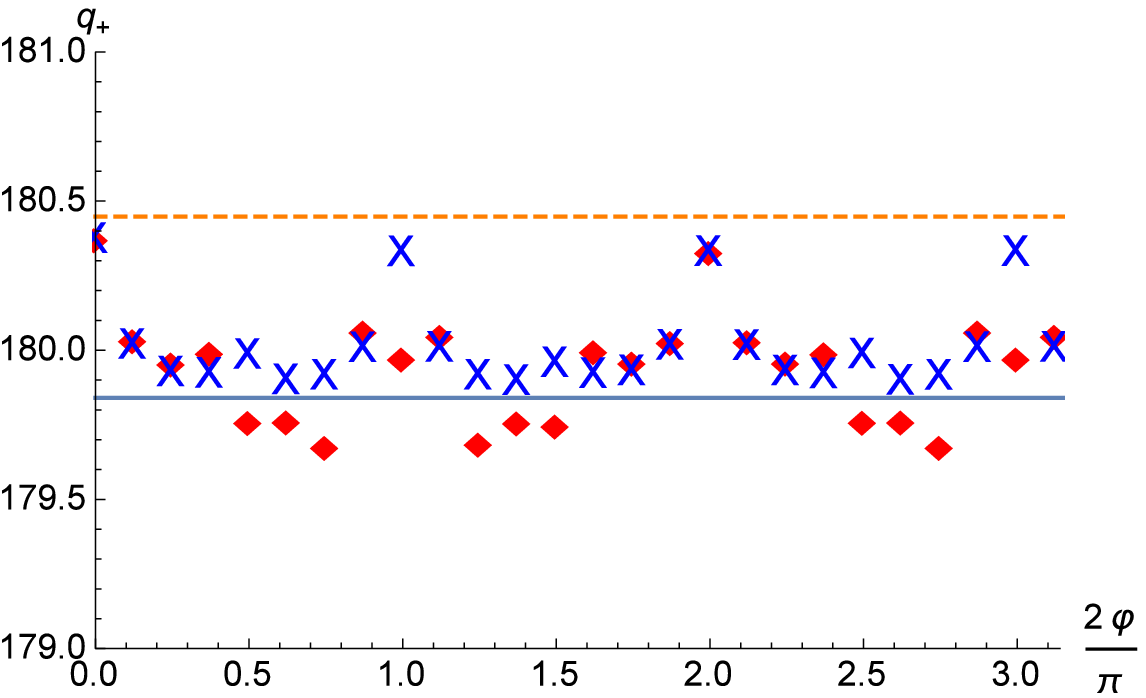}\subcaption{\quad }\label{Fig.5a}
\end{minipage} \hspace{0.01\linewidth}
\begin{minipage}[h]{0.48\linewidth}
\includegraphics[scale=0.6]{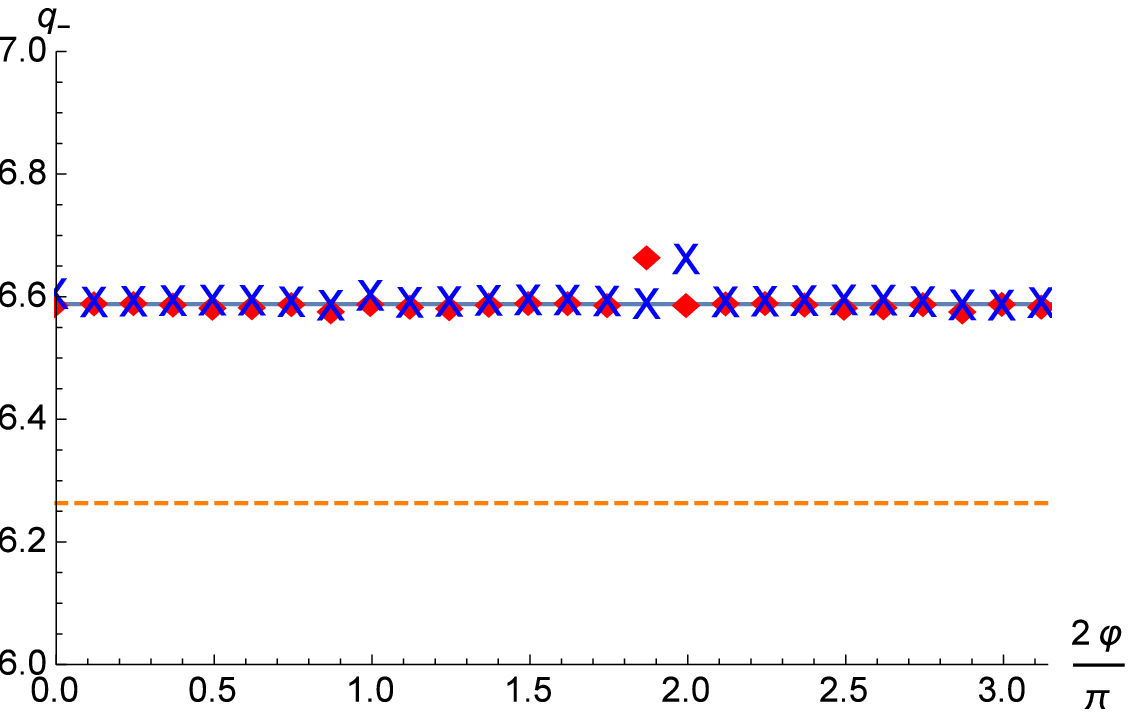}\subcaption{\quad }\label{Fig.5b}
\end{minipage}
\begin{minipage}[h]{0.48\linewidth}
\includegraphics[scale=0.6]{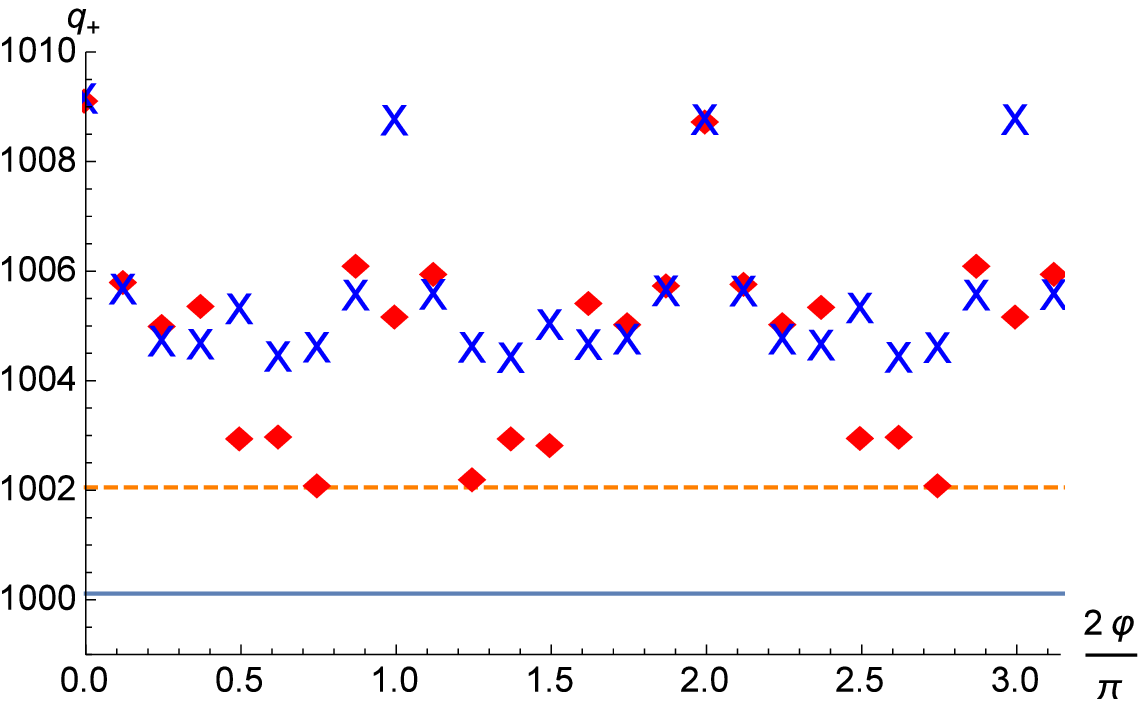}\subcaption{\quad }\label{Fig.6a}
\end{minipage} \hspace{0.01\linewidth}
\begin{minipage}[h]{0.48\linewidth}
\includegraphics[scale=0.6]{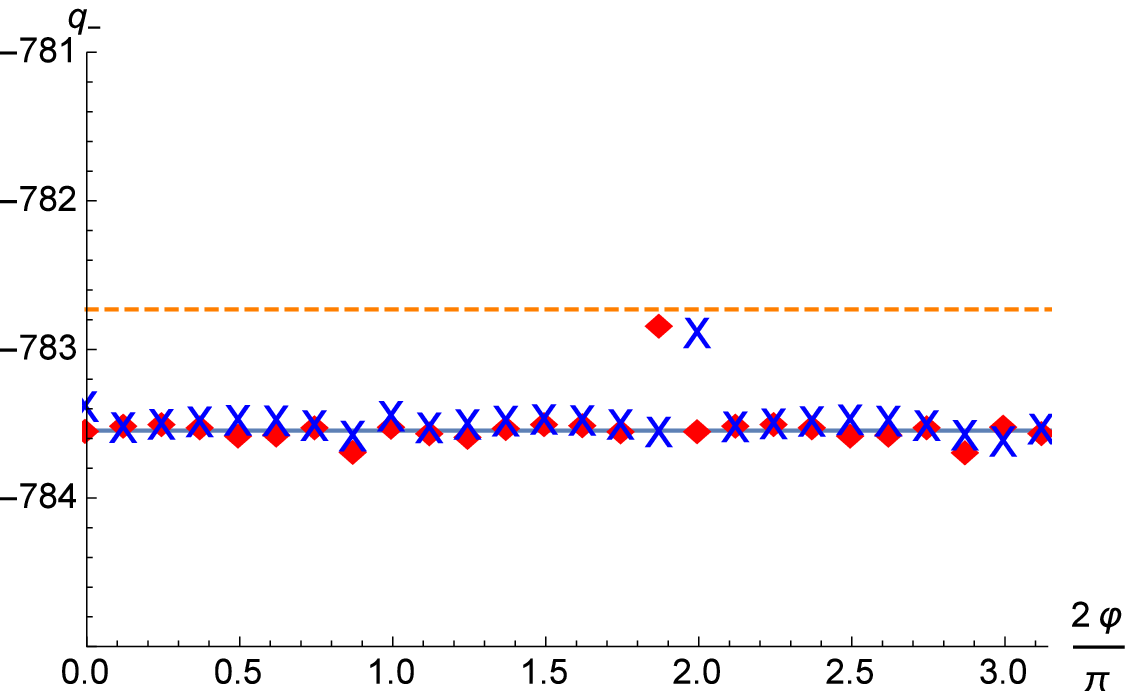}\subcaption{\quad }\label{Fig.6b}
\end{minipage}
\caption{The exact heat flux (solid line), the heat flux evaluated from the simplified problem with the imperfect interface conditions (dashed), and the COMSOL-modelled heat flux (diamond markers for fine mesh, x-markers for normal mesh) along the boundaries, in the cases of monotonic (a-b) and non-monotonic (c-d) $T(r)$ distributions.}
\end{figure}

In finding the precise solution for the original system, we assumed that the temperature and heat flux were constant along the boundaries of the interphases, allowing us to find the analytical solutions. Unsurprisingly, the COMSOL simulations produce slight fluctuations in the value of the heat and temperature fluxes, which is a direct consequence of the non-uniformly distributed mesh in the axisymmetric problem.
These fluctuations in the heat flux, as compared to the values of heat flux evaluated for the original and the simplified problems, can be seen in Fig.~\ref{Fig.5a} -- \ref{Fig.6b}.

Meanwhile, Tables.~\ref{OuterHeatFluxErrors}--\ref{InnerHeatFluxErrors} collect the errors, both absolute and relative, of the heat flux evaluated from the solution of the simplified problem with the imperfect interface replacing the original interphase, and the heat flux in the COMSOL simulations, where `f.m.' stands for `extremely fine mesh' and `n.m.' for the normal mesh. For the sake of brevity, the solution to the simplified problem is denoted by $\widetilde{q}_\pm$ in the tables.

\begin{table}[H]
\begin{center}
\begin{tabular}{l*{11}{c}}
 & $\Delta \tilde{q}_+$ & $\delta \tilde{q}_+$ & $\Delta q_+$, f.m. & $\delta q_+$, f.m. & $\Delta q_+$, n.m. & $\delta q_+$, n.m.\\
\hline
M & 0.6 & 0.003 & 0.5 & 0.003 & 0.528 & 0.003 \\
\hline
N-M & 1.9 & 0.002 & 9.0 & 0.009 & 8.995 & 0.009
\end{tabular}
\caption{Errors for the heat flux at the outer boundary of the interphase.}\label{OuterHeatFluxErrors}
\end{center}
\end{table}
\begin{table}[H]
\begin{center}
\begin{tabular}{l*{11}{c}}
 & $\Delta \tilde{q}_-$ & $\delta \tilde{q}_-$ & $\Delta q_-$, f.m. & $\delta q_-$, f.m. & $\Delta q_-$, n.m. & $\delta q_-$, n.m.\\
\hline
M & 0.3 & 0.052 & 0.56 & 0.09 & 0.07 & 0.01 \\
\hline
N-M & 0.8 & 0.001 & 8.87 & 0.01 & 0.64 & $0.001$
\end{tabular}
\caption{Errors for the heat flux at the inner boundary of the interphase.}\label{InnerHeatFluxErrors}
\end{center}
\end{table}

We note that this part of the analysis is concerned with the accuracy of the FEM computations and not with that of the transmission conditions, as we particularly wish to assess how well both the temperature and flux along the thin interphase are accommodated within the available FEM code. Later, this information allows to make some indirect comparisons without having analytical results.

It appears that the numerical solutions for heat flux produced by COMSOL are more accurate at the inner boundary than at the outer boundary, and that using the normal mesh led to better precision than the fine mesh. However, at the outer boundary, the change of mesh affected the solution but not its precision. This demonstrates an additional limitation of the direct implementation of thin interphases in the FEM model. Even in the case when such computations are practical and tractable, an increase in the number of mesh points does not necessarily lead to an improvement in numerical accuracy.

\subsection{Cases of non-circular interphases}
Having benchmarked the analytical solution via COMSOL computational results and the solution of the simplified problem with transmission conditions for the simple circular case, we consider a more complicated situation, where the thin interphase is only close to being circular.

As it was not possible to find the analytical solution for such a geometry, we use only the results of the COMSOL computations to assess the validity of the transmission conditions in these cases. Thus, we use here the information from the previous section, where the accuracy of the COMSOL computations was thoroughly assessed.
Since the FEM computations are limited by the interphase thicknesses we may model, we provide a verification based on the previous assumption that the thickness of the interphase is equal to $h=0.01$.

All of the physical parameters - except, naturally, the boundaries of the interphase region - are retained from the case of the circular interphase, in which the solution was monotonic. We should note, however, that in this instance the COMSOL computations reveal regions of both monotonic and non-monotonic temperature within the interphase. This effect can be explained by the variation of width in the interphase layer. Nevertheless, the boundaries were defined in such a way that the centre line of the interphase is still the circle $r_0(\phi)=r_0=1$.

We verify the precision of the transmission conditions in the same manner as we did previously for the circular interphase, by substituting the solutions for the temperature and heat flux into the transmission conditions and evaluating the differences between the left and the right sides of the equations thus obtained ($\delta_i^{(m)/(nm)}, i=1,2$). The point values of $T_\pm, q_\pm$ were collected at probe sites situated along the respective boundaries of the interphase, angularly separated by $\pi/16$. The results of this verification are presented in Table.\ref{Verification2}, where the error is the mean relative error among the sets of probe points. They demonstrate that the errors increase for interphases with greater curvature of the boundaries.
\begin{table}[H]
\begin{center}
\begin{tabular}{l*{6}{c}}
relative errors & $\delta_1^{(m)}$ & $\delta_2^{(m)}$ & $\delta_2^{(nm)}$  \\
\hline
$n=0$ & 0.007 & $0.003$ & $0.005$\\
\hline
$n=1$ & 0.06 & 0.013 & 0.006\\
\hline
$n=10$ & 0.08 & 0.005 & 0.012\\
\hline
$n=50$ & 0.27 & 0.13 & 0.16\\
\end{tabular}
\caption{Verification of the transmission conditions for almost circular interphases of different curvature, defined by the parameter $n$ (see (\ref{shapes})).}\label{Verification2}
\end{center}
\end{table}

\begin{figure}[H]
\begin{minipage}[h]{0.48\linewidth}
\includegraphics[scale=0.6]{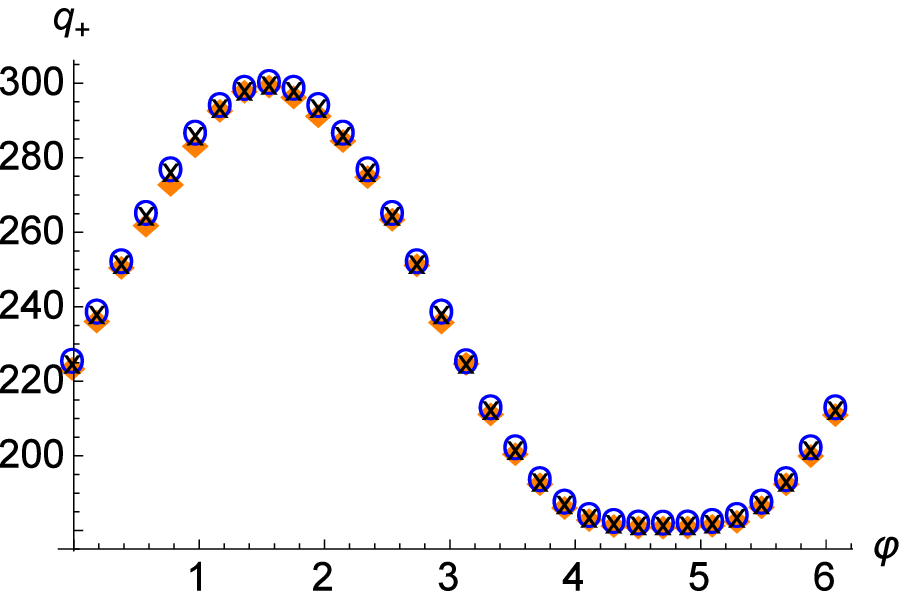}\subcaption{\quad }\label{Fig.7a}
\end{minipage} \hspace{0.01\linewidth}
\begin{minipage}[h]{0.48\linewidth}
\includegraphics[scale=0.6]{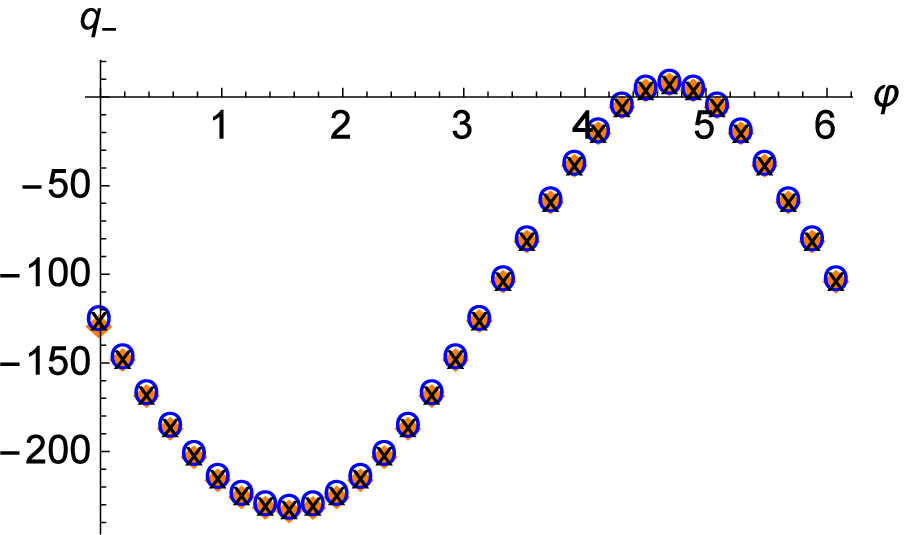}\subcaption{\quad }\label{Fig.7b}
\end{minipage}
\begin{minipage}[h]{0.48\linewidth}
\includegraphics[scale=0.6]{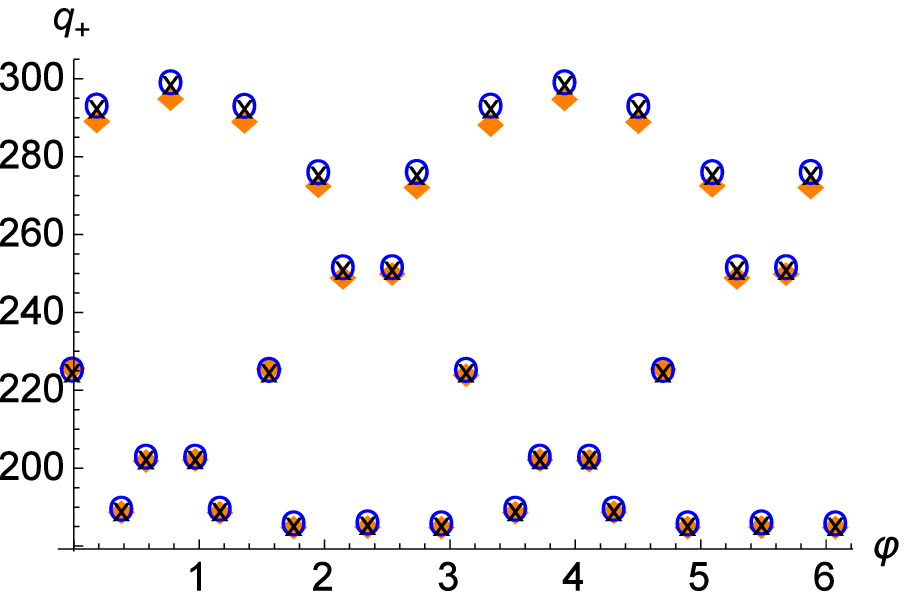}\subcaption{\quad }\label{Fig.8a}
\end{minipage} \hspace{0.01\linewidth}
\begin{minipage}[h]{0.48\linewidth}
\includegraphics[scale=0.6]{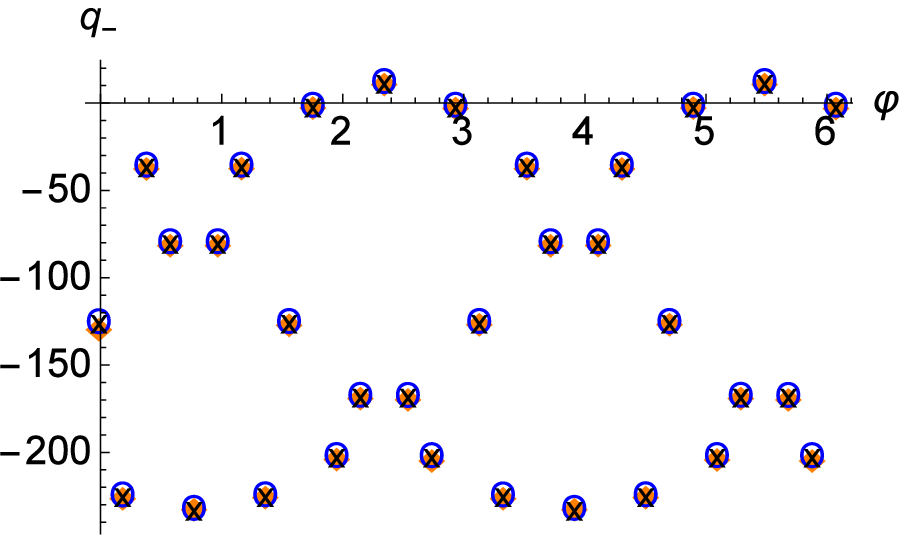}\subcaption{\quad }\label{Fig.8b}
\end{minipage}
\begin{minipage}[h]{0.48\linewidth}
\includegraphics[scale=0.6]{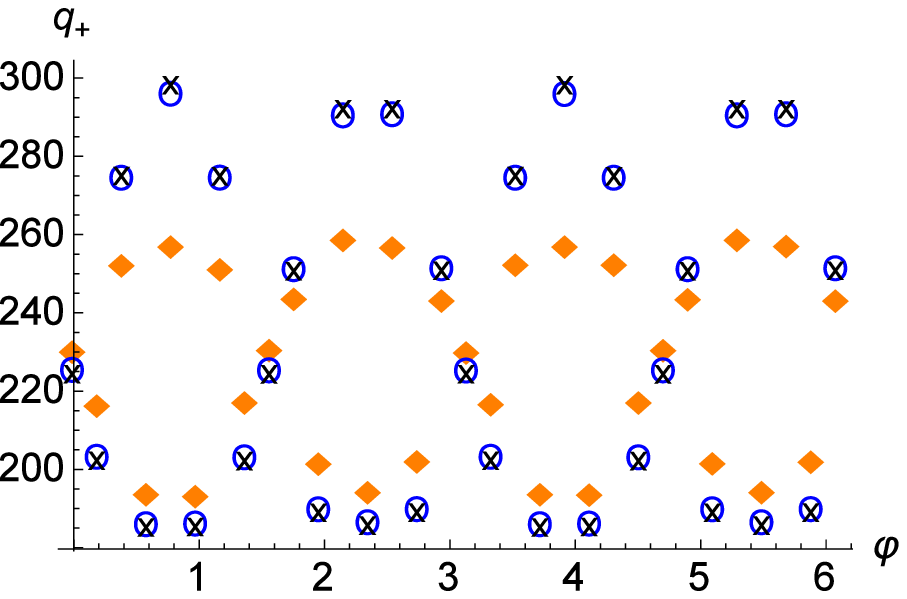}\subcaption{\quad }\label{Fig.9a}
\end{minipage} \hspace{0.01\linewidth}
\begin{minipage}[h]{0.48\linewidth}
\includegraphics[scale=0.6]{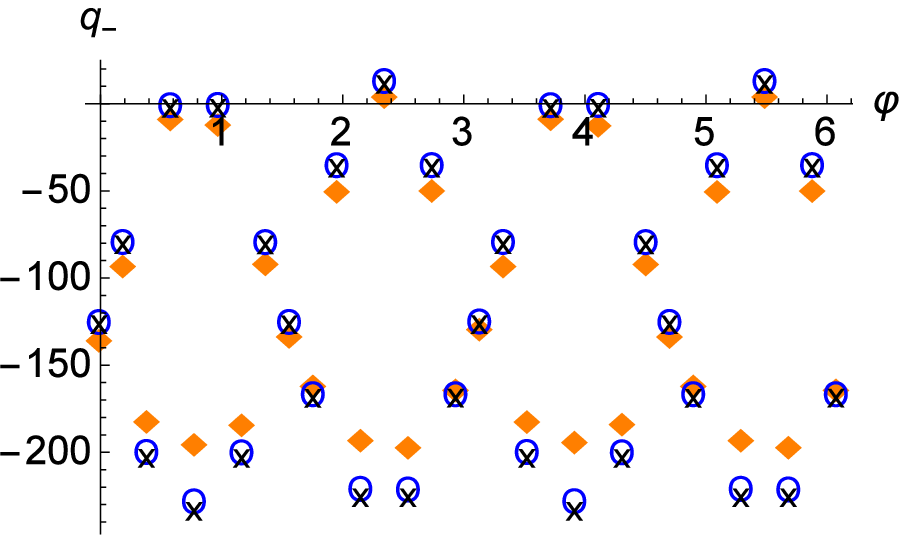}\subcaption{\quad }\label{Fig.9b}
\end{minipage}
\caption{Heat flux along boundaries of different curvature: diamond markers for COMSOL values, o-shaped and x-shaped for the values evaluated via the two sets (with and without assumed monotonicity) of transmission conditions. Results shown for the boundaries with $n=1$ (a-b), $n=10$ (c-d), $n=50$ (e-f)).}
\end{figure}

Another (indirect) way to assess the transmission conditions for interphases of different curvature is to use them directly to evaluate the heat fluxes, $q_+$ and $q_-$, from both sides of the interphase using the temperatures $T_+$ and $T_-$ computed in COMSOL. We do this by substituting the numerical solution for temperature into the transmission conditions and evaluating the heat flux from the resulting systems of equations. The results are shown in  Fig.~\ref{Fig.7a}--\ref{Fig.9b}.
Here we take advantage of the fact that the temperature is computed in FEM much more accurately than the heat flux. We then compare those values with the fluxes evaluated by COMSOL.
As can be seen from the figures, the solutions calculated from the two sets of transmission conditions, those with assumptions of monotonicity and without, give similar results. However, for the interphase with the greatest curvature ($n=50$), they significantly deviate from one other. We note that in the case $n=10$, although the curvature is essentially larger than in the case $n=1$, the transmission conditions still provide very good accuracy.

The absolute and relative errors in each case are collected in Table \ref{HeatFluxErrors}. We can see that for boundaries characterized by a smaller curvature ($n=1$, $n=10$), the errors are of the same order and are still reasonable, but have lost an order of accuracy in comparison with the circular interphase. However, in the case of a greater curvature ($n=50$), a drastic loss of accuracy is evident.

\begin{table}[H]
\begin{center}
\begin{tabular}{l*{8}{c}}
 & $\Delta q_+^{(m)}$ & $\delta q_+^{(m)}$ & $\Delta q_-^{(m)}$ & $\delta q_-^{(m)}$ & $\Delta q_+^{(nm)}$ & $\delta q_+^{(nm)}$ & $\Delta q_-^{(nm)}$ & $\delta q_-^{(nm)}$ \\
\hline
$n=0$ & 0.74 & 0.004 & 0.44 & 0.066 & 0.74 & 0.004 & 0.44 & 0.066\\
\hline
$n=1$ & 3.27 & 0.012 & 2.86 & 0.095 & 3.23 & 0.012 & 3.0 & 0.095\\
\hline
$n=10$ & 4.11 & 0.014 & 2.96 & 0.091 & 4.06 & 0.014 & 3.1 & 0.091\\
\hline
$n=50$ & 38.4 & 0.149 & 35.6 & 4.658 & 41.3 & 0.161 & 39.5 & 4.716\\
\end{tabular}
\caption{Accuracy of solutions for heat flux, evaluated via the transmission conditions for almost circular interphases of different curvature.}\label{HeatFluxErrors}
\end{center}
\end{table}

\section{Conclusions and discussion}
In this work we have evaluated the transmission conditions for nonlinear thin reactive interphases. Both the general case and a special case, where the temperature distribution inside the interphase was monotonic, were considered in detail. The conditions were evaluated and verified for several classes of the material parameters - conductivities of the interphase material and possible sources (sinks) $Q$.

We have shown, by dedicated numerical and analytical experiments, that the conditions work satisfactorily and give a relatively small error, on the order at least of $O(h^2)$ as expected. Interestingly, even if the temperature was not monotonic within the interphase, the transmission conditions evaluated under opposing assumptions may sometimes provide valuable results. Clearly, the general transmission conditions should be used in further analyses if the temperature field is not known \textit{a priori}.

We have also shown a reduction in the accuracy of the conditions evaluated in this study when the main assumption used in their evaluation, that of small curvature, was violated.

We have restricted ourselves to the case where the centre line of the interphase coincides with a circle. This slightly simplifies the analysis and allows more elegant analytical results in evaluating the transmission conditions. However, the general case ($r_0(\phi) \neq 0$) can also be considered if the main assumption on the curvature is still valid.

\section{Acknowledgements}
\noindent
DA and AZ acknowledge the support from FP7 Marie Curie IAPP project IN-TERCER2 (PIAP-GA-2011-286110-INTERCER2) and while WM was supported by the IAPP project PARM2 (PIAP-GA-2012-284544-PARM2).

DA is thankful to the Department of Civil, Environmental and Mechanical Engineering (Solid and Structural Mechanics Group) of the University of Trento (Italy) for assistance with the COMSOL Multiphysics$^{\circledR}$ during her secondment.

\bibliography{Bibliography}{}

\appendix

\section[Evaluation]{Evaluation of transmission conditions for some specific cases}
\subsection{Evaluation of the transmission conditions in the linear case}
By intergrating (\ref{BVP}), and introducing an auxiliary function $\Phi(\xi, \phi, t)$, we obtain
\begin{equation}
\label{SecondDiffEquation1}
\frac{\widetilde{k}}{\widetilde{h}} \frac{\partial T}{\partial \xi} = - \widetilde{h}(\phi) \Phi(\xi, \phi, t) + C_0(\phi, t),
\end{equation}
where
\begin{equation}\label{PhiFunction1}
\Phi(\xi, \phi, t) = \int_{-\frac{1}{2}}^\xi \widetilde{Q} (z, \phi, t) dz.
\end{equation}

It follows from substituting $\xi = -\frac{1}{2}$ into (\ref{SecondDiffEquation1}) that
$
C_0(\phi, t) = - q_{-}, \label{FirstConstant1},
$
while substituting $\xi = \frac{1}{2}$ leads to the first transmission condition (\ref{Conditions_Case_1})$_1$
\begin{equation}\label{FirstCondition1}
q_{+} - q_{-} = \widetilde{h}(\phi) \Phi\Big(\frac{1}{2}, \phi, t\Big).
\end{equation}

To evaluate the second transmission condition, it is sufficient to integrate the equation (\ref{SecondDiffEquation1}), taking into account the value of $C_0(\phi, t)$, to obtain
\begin{equation}
\label{ThirdEquation1}
T(\xi,\phi,t) = - \widetilde{h}^2(\phi) \Psi(\xi, \phi, t) - \tilde h(\phi) q_- \Xi(\xi, \phi, t) + C_1(\phi,t).
\end{equation}
Here, we have introduced two auxiliary functions
\begin{equation}
\Psi(\xi, \phi, t) = \int_{-\frac{1}{2}}^\xi \frac{\Phi(z, \phi, t)}{\widetilde{k}(z, \phi, t)} dz, \quad
\Xi(\xi, \phi, t) = \int_{-\frac{1}{2}}^\xi \frac{dz}{\widetilde{k}(z, \phi, t)}.
\label{AuxiliaryFunctions1}
\end{equation}
The integration constant
\[
C_1(\phi,t) = T\Big(-\frac{1}{2},\phi,t\Big),
\]
is found from (\ref{ThirdEquation1}) by taking $\xi = -1/2$.

Finally, by setting in (\ref{ThirdEquation1}) $\xi = 1/2$, the second condition (\ref{Conditions_Case_1})$_2$ is obtained
\begin{equation}\label{SecondCondition1}
T_+ - T_- =  - \widetilde{h}(\phi) q_- \Xi\Big(\frac{1}{2}, \phi, t\Big)-\widetilde{h}^2(\phi) \Psi\Big(\frac{1}{2}, \phi, t\Big).
\end{equation}

\subsubsection*{Special case (1a): $\widetilde{Q} = \widetilde{Q}_0$, $\widetilde{k} = \widetilde{k}_0$}
\label{SC_1a}
If we suppose that both the thermal source and thermal conductivity are independent of the temperature and the coordinates, the transmission conditions (\ref{Conditions_Case_1}) are easily obtained by the appropriate substitution into (\ref{FirstCondition1}) and (\ref{SecondCondition1}) as
\begin{equation}
\begin{array}{rcl}
q_+- q_- &=& \widetilde{h}(\phi)\tilde Q_0, \quad \\[2mm]
T_+- T_- &=& -\frac{\widetilde h(\phi)}{2\widetilde k_0}(q_+ + q_-).
\end{array}
\label{Conditions_Case_1a}
\end{equation}

When considering the linear case, no additional assumptions on the behaviour of the temperature within the interphase were necessary. However, for the other cases, different transmission conditions can be obtained depending on whether $T(r)$ is asssumed to be monotonic or not.

\subsection{Evaluation of the transmission conditions in case of the thermal conductivity being a function of temperature}
\subsubsection*{Transmission conditions for monotonic temperature distribution in the interphase}
As was mentioned before, the first transmission condition is the same as in the previous case. To obtain the second condition, we need to integrate an equation analogous to (\ref{SecondDiffEquation1}),
\begin{equation}
\label{SecondDiffEquation2}
\widetilde{k}(T)T' = - \widetilde{h}^2\frac{\Phi(\xi, \phi, t)}{m(\xi, \phi, t)} - q_- \frac{\widetilde{h}^2}{m(\xi, \phi, t)}.
\end{equation}

The assumption of monotonicity of the temperature plays an important role at this step, as it allows us to change the integration variable in the left-hand side of the equation. As a result, we obtain
\begin{equation}
\Theta(T,T_-, \phi, t) = -\widetilde{h}^2(\phi)\Psi(\xi, \phi, t) - q_-\widetilde{h}(\phi)\Xi(\xi, \phi, t) + C_1(\phi,t),
\end{equation}
where the introduced auxiliary functions are
\begin{equation}
\Theta = \int_{T_-}^T \widetilde{k}(z, \phi, t) dz,  \quad \Psi= \int_{-\frac{1}{2}}^\xi \frac{\Phi(z, \phi, t)}{m(z, \phi, t)} dz, \quad
\Xi= \int_{-\frac{1}{2}}^\xi \frac{dz}{m(z, \phi, t)}.
\label{AuxiliaryFunctions2}
\end{equation}
$C_1(\phi,t)$ evidently vanishes (verified by substitution of $\xi = -1/2$), and the second transmission condition takes its final form
\begin{equation}\label{SecondCondition2}
\Theta(T_+,T_-, \phi, t) = -\widetilde{h}^2(\phi)\Psi\Big(\frac{1}{2}, \phi, t\Big) - q_- \widetilde{h}(\phi)\Xi\Big(\frac{1}{2}, \phi, t\Big).
\end{equation}

\subsubsection*{Special case (2a): $\widetilde{Q} = \widetilde{Q}_0, \quad \widetilde{k} = \widetilde{k}_0$}
Substituting constant values into (\ref{AuxiliaryFunctions2}) and (\ref{SecondCondition2}) gives the same result as in the previously considered \ref{SC_1a}.

\subsubsection*{Special case (2b): $\widetilde{Q} = \widetilde{Q}_0, \quad \widetilde{k} = \widetilde{k}_0+\widetilde{k}_1\tilde T$}\label{SC_2b}
In this instance, the first transmission condition remains the same as for the previously mentioned special cases. The second transmission condition (\ref{SecondCondition2}) takes the form
\begin{equation}\label{Conditions_Case_2b}
\widetilde{k}_0(T_+ - T_-) + \frac{\widetilde{k}_1}{2}(T_+^2 - T_-^2) = -\frac{q_+^2 - q_-^2}{2 \widetilde{Q}_0}.
\end{equation}

\subsubsection*{Transmission conditions without additional assumptions on monotonicity}
Let us make an important observation in the case of non-monotonic behaviour of the solution within the interphase in the direction perpendicular to the interphase surfaces, for some values $\phi$ and $t$. In this instance, there should exist  $\xi = \xi_*(\phi,t) \in (-1/2, 1/2)$, where $T$ takes its extremal value $T_*$ \big($T'_\xi(\xi_*) = 0$\big).

Generally speaking, there may exist a few extremal points. To rule out such a possibility, we assume in this paper that the thermal load $Q$ within the interphase represents only a source ($Q\ge0$) or a sink ($Q\le0$). It is then clear from physical arguments that for any chosen values $\phi$ and $t$, there may exist only one extremal value of the temperature $T_*=T_*(\phi,t)$ at its respective point inside the interphase $\xi_* = \xi_*(\phi,t)$. As a result, the interval $(-1/2,1/2)$ splits into two, $(-1/2, \xi_*)$ and $(\xi_*, 1/2)$. Note that equation (\ref{SecondDiffEquation2}) can be written in each of the subdomains in an equivalent form,
\begin{equation}
\widetilde{k}(T, \phi, t) m(\xi, \phi, t)T' = -\widetilde{h}^2(\phi)\Phi_\pm(\xi,\phi,t) - \widetilde{h}(\phi)q_\pm, \label{Equations2}
\end{equation}
where the auxiliary functions are defined by
\begin{equation}\label{PhiFunctions2}
\Phi_\pm(\xi, \phi, t) = \int_{\pm\frac{1}{2}}^{\xi} \widetilde{Q}(z,\phi, t) dz.
\end{equation}

Considering both expressions at the extremal point $\xi_*$, and bearing in mind that $T'_* = 0$, we find
\begin{equation}\label{PreFirstCondition2}
\widetilde{h}(\phi)\Phi_\pm(\xi_*,\phi,t) = -q_\pm,
\end{equation}
which gives the first transmission condition (\ref{Conditions_Case_2_NM})$_1$ in an equivalent form,
\begin{equation}\label{FirstNMCondition}
\Phi_+^{-1}\Big({-q_+}\big/{\,\widetilde{h}(\phi)}\Big) = \Phi_-^{-1}\Big({-q_-}\big/{\,\widetilde{h}(\phi)}\Big).
\end{equation}

This also provides the formula for finding the point $\xi_*$,
\begin{equation}
\xi_* = \Phi_+^{-1} \left({-q_+}\big/{\,\widetilde{h}(\phi)} \right) \quad \text{or} \quad  \xi_* = \Phi_-^{-1} \left({-q_-}\big/{\,\widetilde{h}(\phi)}\right).
\end{equation}
To evaluate the second transmission condition, we integrate (\ref{Equations2}) to obtain
\begin{equation}\label{NMAsymptoticPrediction}
\Theta_\pm(T, \phi, t) = -\widetilde{h}^2(\phi)\Psi_\pm(\xi, \phi, t) - \widetilde{h}(\phi) q_\pm \Xi_\pm(\xi, \phi, t).
\end{equation}
where
\begin{equation}
\Theta_\pm= \int_{T_\pm}^T \widetilde{k}(z, \phi, t) dz,  \quad
\Psi_\pm= \int_{\pm\frac{1}{2}}^\xi \frac{\Phi_\pm(z, \phi, t)}{m(z, \phi, t)} dz, \quad
\Xi_\pm= \int_{\pm\frac{1}{2}}^\xi \frac{dz}{m(z, \phi, t)}.
\label{AuxiliaryFunctions2_NM}
\end{equation}

We note that the functions $\Theta_\pm$ are monotonic with respect to the temperature $T$, and thus that there exist inverse functions to $\Theta_\pm(T, \phi, t)$ with respect to $T$.
Substituting the values of the parameters corresponding to the common point $\xi=\xi_*$ into such representations of temperature gives the value of $T_*$, and therefore leads to the second transmission condition
\begin{equation}
\begin{array}{r}
\Theta_+^{-1}\Big(-q_+ \widetilde{h}(\phi)\Xi_+(\xi_*, \phi, t) - \widetilde{h}^2(\phi)\Psi_+(\xi_*, \phi, t)\Big) = \\[2mm]
\Theta_-^{-1}\Big(-q_- \widetilde{h}(\phi)\Xi_-(\xi_*, \phi, t) - \widetilde{h}^2(\phi)\Psi_-(\xi_*, \phi, t)\Big).
\end{array}
\label{SecondCondition2_NM}
\end{equation}

\textit{Remark.} The cases where $q_\pm=0$ imply monotonic temperature distribution, with $T$ reaching its extremal values at $\pm 1/2$. This is evident from substitution of zero heat flux into (\ref{PreFirstCondition2}), that gives
$
\Phi_+(\xi_*, \phi, t) = 0.
$

We note that in the two special cases discussed above under the assumption of monotonic temperature distribution ($\widetilde{Q} = \widetilde{Q}_0,  \widetilde{k} = \widetilde{k}_0$ and $\widetilde{Q} = \widetilde{Q}_0, \quad \widetilde{k} = \widetilde{k}_0+\widetilde{k}_1\widetilde{T}$) the transmission conditions have the same form (\ref{Conditions_Case_1a}) or (\ref{Conditions_Case_2b}) respectively, also when this assumption is discarded.

\subsection{Evaluation of the transmission conditions for temperature-dependent sources and material properties}
In this case we can again obtain the transmission conditions by solving (\ref{BVP}), and introducing auxiliary functions at each step. The functions used in this case are
\begin{equation}
\Phi_\pm(T) = \int_{T_\pm}^T  \widetilde{k}(z) \widetilde{Q}(z) dz,  \quad  \Upsilon_\pm(T) = \int^T_{T_\pm} \frac{\widetilde{k}(z)}{\sqrt{(q_\pm)^2 - 2 \Phi_\pm(z)}}dz,
\label{AuxiliaryFunctions3}
\end{equation}
where using the function $\Upsilon_\pm(T)$ leads to
\begin{equation}
(q_+)^2 > 2 \Phi_+(T), \quad (q_-)^2 > 2 \Phi_-(T).
\end{equation}

It is important to emphasize at this point that
\begin{equation}
\int_{T_{\pm}}^T  \widetilde{k}(z) \widetilde{Q}(z)dz \leq 0,
\end{equation}
or, equivalently, that the following should always be satisfied,
\begin{equation}
(T_+ - T_-) \widetilde{Q}(z) < 0.
\end{equation}

When we additionally assume that $T(r)$ is monotonic, the transmission conditions obtained are
\begin{eqnarray}\label{FirstCondition3}
(q_+)^2 - (q_-)^2 &=& \pm 2 \Phi_\pm(T_\mp),\\
\label{SecondCondition3}
\Upsilon_\pm(T_\mp) &=& \pm\widetilde{h}(\phi),
\end{eqnarray}
where the corresponding set of conditions, with $\Phi_+, \Upsilon_+$ or $\Phi_-, \Upsilon_-$ should be used when $q_+ \neq 0$ or $q_- \neq 0$.

By analogy, we obtained the following transmission conditions without additional assumptions of monotonicity,
\begin{eqnarray}\label{FirstCondition3_NM}
\Phi_+^{-1}\Big(\frac{q_+^2}{2}\Big) &=& \Phi_-^{-1}\Big(\frac{q_-^2}{2}\Big), \\
\label{SecondCondition3b_NM}
\Upsilon_+(T_*) - \Upsilon_-(T_*) &=& \widetilde{h}(\phi).
\end{eqnarray}

Either the left hand or the right hand side of (\ref{FirstCondition3_NM}) can be used to evaluate the maximum/minimum value of the temperature, $T_*$. At the same time, the point $\xi_*$ at which the temperature reaches this value, can be found from the following formula, obtained in the process of deriving the second transmission condition (\ref{SecondCondition3b_NM}),
\begin{equation}\label{ExtremumPoint}
\xi_* = \frac{1}{2} - \frac{\Upsilon_-(T_*)}{\widetilde{h}(\phi)} \quad \text{or} \quad  \xi_* = -\frac{1}{2} - \frac{\Upsilon_+(T_*)}{\widetilde{h}(\phi)}.
\end{equation}

\subsubsection*{Special case (3a): $\widetilde{Q}(T) = \widetilde{Q}_0, \quad \widetilde{k} = \widetilde{k}_0$}
If both mentioned parameters are constant, the transmission conditions take the forms that, after some simplifications, can be transformed into the same as in 
(\ref{Conditions_Case_1a}).

\subsubsection*{Special case (3b): $\widetilde{Q} = \widetilde{Q}_0, \widetilde{k} = \widetilde{k}_0 + \widetilde{k}_1 T$}
Both sets of transmission conditions - with and without assuming monotonicity of the temperature function - transform in this instance to a form equivalent to the conditions described in the special case (2b) in (\ref{Conditions_Case_2b}).

\subsubsection*{Special case (3c): $\widetilde{Q}(T) = \widetilde{Q}_0(\tilde T + \beta), \quad \widetilde{k} = \widetilde{k}_0$}
For this special case, the form of the second transmission condition depends on whether we are considering a case of heat source ($\widetilde{Q}_0 > 0$) or heat sink ($\widetilde{Q}_0 < 0$). As $\widetilde{Q}_0 \neq 0$, we shall not consider cases of negative $\beta$. Substituting these expressions into (\ref{AuxiliaryFunctions3}) leads to the conditions (\ref{FirstCondition3}) and (\ref{SecondCondition3}) which, after a series of simplifying transformations, take the forms
\begin{equation}
\begin{array}{c}
\widetilde{k}_0 \widetilde{Q}_0 (2\beta(T_- - T_+) + (T_-^2 - T_+^2)) \quad = \quad q_+^2 - q_-^2, \\[2mm]

\frac{q_-^2 + \widetilde{k}_0 \widetilde{Q}_0 (T_- + \beta)^2}{\sqrt{\widetilde{k}_0 \widetilde{Q}_0}} \sin\Big(-\widetilde{h}(\phi)\sqrt{\frac{\widetilde{Q}_0}{\widetilde{k}_0}}\Big) = \\[2mm]
(T_+ + \beta)\vert q_- \vert - (T_- + \beta)\sqrt{q_-^2 + \widetilde{k}_0 \widetilde{Q}_0 ((T_- + \beta)^2 - (T_+ + \beta)^2)},
\end{array}
\label{Conditions_Case_3b}
\end{equation}
or, in the case of heat sink,
\begin{align}
T_+ + \beta + \sqrt{\frac{q_-^2}{-\widetilde{k}_0 \widetilde{Q}_0} + ((T_- + \beta)^2 - (T_+ + \beta)^2)} =
\frac{\exp\Big(\widetilde{h}(\phi)\sqrt{-\frac{\widetilde{Q}_0}{\widetilde{k}_0}}\Big)}{\sqrt{\widetilde{k}_0 \widetilde{Q}_0}}.
\end{align}

For a non-monotonic temperature distribution within the interphase, the first transmission condition is found to be the same as for the monotonic distribution, whilst the second is, in the case of a heat source,
\begin{equation}
\begin{array}{c}
\sqrt{\widetilde{k}_0 \widetilde{Q}_0} (\vert q_- \vert (T_+ + \beta) - \vert q_+ \vert (T_- + \beta)) = \\ \sqrt{(q_+^2 + {\widetilde{k}_0 \widetilde{Q}_0} (T_+ + \beta)^2)(q_-^2 + {\widetilde{k}_0 \widetilde{Q}_0} (T_- + \beta)^2)} \sin\Big(-\widetilde{h}(\phi)\sqrt{\frac{\widetilde{Q}_0}{\widetilde{k}_0}}\Big),
\end{array}
\label{Conditions_Case_3b_NM}
\end{equation}
or, for $\widetilde{Q}_0 < 0$,
\begin{equation}
T_- + \beta - \frac{q_-^2}{-\widetilde{k}_0 \widetilde{Q}_0} = \Big(T_+ + \beta - \frac{q_+^2}{-\widetilde{k}_0 \widetilde{Q}_0}\Big) \exp\Big(2 \widetilde{h}(\phi)\sqrt{-\frac{\widetilde{Q}_0}{\widetilde{k}_0}}\Big).
\end{equation}

\section{Explicit formulae of the benchmark solutions}
\subsection[A2.2]{Monotonic temperature distribution within the interphase}
The exact solutions to the original problem with a thin interphase, after the substitution of the numerical values of the parameters into the previously evaluated formulae, are, in the respective three regions
\begin{eqnarray*}
T_-(r) &=& 299.981 - 0.028 \log r, \quad T_+(r) =295.311 - 0.766 \log r,\\[2mm]
T(r) &=& -50 -1374.13 J_0(15.8114 r) + 1075.84 Y_0(15.8114 r).
\end{eqnarray*}
Here, $J_0, Y_0$ are the Bessel functions of the first and second kind.

The solutions to the simplified problem with the imperfect interface replacing the interphase, that approximate the solutions to the original problem, are
\begin{eqnarray*}
T_2(r) &=& 299.982 - 0.026 \log r, \quad T_1(r) = 295.312 - 0.769 \log r,\\[2mm]
\widetilde{T}(r) &=& -50 + 345.304 \cos(-15.811 r + 15.97) - 57.063 \sin(-15.811 r + 15.97).
\end{eqnarray*}

\subsection[A2.2]{Non-monotonic temperature distribution within the interphase}
The temperatures in the three regions, after substitution of the numerical solutions into the previously evaluated exact solutions to the original problem with a thin interphase, are
\begin{eqnarray*}
T_-(r) &=& 302.292 + 3.306 \log r, \quad T_+(r) = 296.728 - 4.262 \log r,\\[2mm]
T(r) &=& -50 + 2174.96 J_0(50 r) - 2354.58 Y_0(50 r).
\end{eqnarray*}

The solutions to the simplified problem with the transmission conditions, that approximate the solutions to the original problem, are
\begin{eqnarray*}
T_2(r) &=& 302.289 + 3.303 \log r,\quad T_1(r) = 296.731 - 4.27 \log r,\\[2mm]
\widetilde{T}(r) &=& \left\{\begin{array}{lr}
- 50 + 352.289 \cos(50.5 - 50 r) +78.273 \sin(50.5 - 50 r), \\[2mm]
- 50 + 346.689 \cos(50.5 - 50 r) -100.205 \sin(50.5 - 50 r),
\end{array}\right.
\end{eqnarray*}
where the respective expressions for $\widetilde{T}(r)$, the solution within the interphase region, correspond to the intervals into which this region was split, $r_- \leq r \leq r_*$ and $ r_* \leq r \leq r_+$. Here
$r_*$ is the extremum point, found from the above mentioned formula (\ref{ExtremumPoint}). We should note that using them in the monotonic case would give a point closest to the interphase, though not within it, where the extended $\widetilde{T}(r)$ has a local extremum.
For example, for the numerical parameters previously considered, if we extend $T(r)$, which is monotonic within the interphase, to a wider domain, this extended function would reach its maximum at a point $\xi_* < 1$, which is approximately equal to 0.9999. Meanwhile, formula (\ref{FirstCondition3_NM}) would also yield a maximum point at approximately 0.9999.

In the case of non-monotonic temperatures, $T(r)$ has maximum at $1.004$. Its asymptotic approximation $\widetilde{T}$ has a maximum at approximately 1.004.

\section{Formulae for verification of transmission conditions}
For $i=1,2$
\begin{equation*}
\delta_i^{(m)} = \frac{\vert LHS_i^{(m)} - RHS_i^{(m)} \vert}{\min(\vert LHS_i^{(m)} \vert, \vert RHS_i^{(m)} \vert)}, \quad
\delta_i^{(nm)} = \frac{\vert LHS_i^{(nm)} - RHS_i^{(nm)} \vert}{\min(\vert LHS_i^{(nm)} \vert, \vert RHS_i^{(nm)} \vert)},
\end{equation*}
where (from (\ref{Conditions_Case_3b}) and (\ref{Conditions_Case_3b_NM}))
\begin{eqnarray*}
LHS_1^{(m)} &=& LHS_1^{(nm)} = \widetilde{k}_0 \widetilde{Q}_0 (n_\xi^0(\phi))^2 (2\beta(T_- - T_+) + (T_-^2 - T_+^2)), \\
RHS_1^{(m)} &=& RHS_1^{(nm)} = q_+^2 - q_-^2, \\
LHS_2^{(m)} &=& \frac{q_-^2 + \widetilde{k}_0 \widetilde{Q}_0 (n_\xi^0(\phi))^2 (T_- + \beta)^2}{n_\xi^0(\phi) \sqrt{\widetilde{k}_0 \widetilde{Q}_0}} \sin\Big(-\widetilde{h}(\phi)\sqrt{\frac{\widetilde{Q}_0}{\widetilde{k}_0}}\Big), \\
RHS_2^{(m)} &=& (T_+ + \beta)\vert q_- \vert - (T_- + \beta)\sqrt{q_-^2 + \widetilde{k}_0 \widetilde{Q}_0 (n_\xi^0(\phi))^2 ((T_- + \beta)^2 - (T_+ + \beta)^2)}, \\
LHS_2^{(nm)} &=& \sqrt{\widetilde{k}_0 \widetilde{Q}_0} n_\xi^0(\phi)(\vert q_- \vert (T_+ + \beta) - \vert q_+ \vert (T_- + \beta)), \\
RHS_2^{(nm)} &=&  \sin\Big(-\widetilde{h}(\phi)\sqrt{\frac{\widetilde{Q}_0}{\widetilde{k}_0}}\Big) \\
&*& \sqrt{(q_+^2 + {\widetilde{k}_0 \widetilde{Q}_0} (n_\xi^0(\phi))^2 (T_+ + \beta)^2)(q_-^2 + {\widetilde{k}_0 \widetilde{Q}_0} (n_\xi^0(\phi))^2 (T_- + \beta)^2)}.
\end{eqnarray*}

\end{document}